\documentclass[preprint,review,12pt,leqno]{elsarticle}

\usepackage{float}
\usepackage{graphicx}
\usepackage{multirow}
\usepackage{pdflscape}
\usepackage{booktabs}
\usepackage{color,amsmath,amssymb}
\usepackage{dcolumn}
\usepackage{geometry}
\usepackage{pifont}
\usepackage{fleqn}
\usepackage{hyperref}
\usepackage{url}
\usepackage[T1]{fontenc}
\usepackage{ae,aecompl}
\usepackage{pdflscape}
\usepackage[normalem]{ulem}
\usepackage{array}
\usepackage{epsf}
\usepackage{epstopdf}
\usepackage{amssymb}
\usepackage{longtable}
\usepackage{epsfig}
\usepackage{gensymb}
\usepackage{makecell}
\newcommand{\specialcell}[2][c]{\begin{tabular}[#1]{@{}c@{}}#2\end{tabular}}

\newcommand{\hhs}{H$_2$$^{32}$S}
\newcommand{\cm}{cm$^{-1}$}
\newcommand{\Marvel}{{\sc Marvel}}

\usepackage{units}

\usepackage[varg]{txfonts}
\usepackage{verbatim}

\begin{document}
\sloppy
\title{\Marvel\ analysis of the measured high-resolution rovibrational spectra of \hhs}

\author{Katy L. Chubb$^a$\footnote{email: katy.chubb.14@ucl.ac.uk}}
\author {Olga Naumenko$^b$}
\author{Stefan Keely$^c$, Sebestiano Bartolotto$^c$, Skye Macdonald$^c$, Mahmoud Mukhtar$^c$, Andrey Grachov$^c$, Joe White$^c$, Eden Coleman$^c$}
\author{Anwen Liu$^d$, Alexander Z.  Fazliev$^b$, Elena R. Polovtseva$^b$, Veli-Matti Horneman$^e$, Alain  Campargue$^f$}
\author{Tibor Furtenbacher$^g$, Attila G. Cs\'asz\'ar$^g$} 
\author{Sergei N. Yurchenko$^a$, Jonathan Tennyson$^a$\footnote{email: j.tennyson@ucl.ac.uk}}
\address{$^a$ Department of Physics and Astronomy, University College London, London, WC1E 6BT, UK}
\address{$^b$ V. E. Zuev Institute of Atmospheric Optics, Siberian Branch Russian Academy of Sciences, Tomsk, Russia}
\address{$^c$ Highams Park School, Handsworth Avenue, Highams Park, London, E4 9PJ, UK}
\address{$^d$ Hefei National Laboratory for Physical Sciences at Microscale,  University of Science and Technology of China, Hefei 230026, China}
\address{$^e$ Department of Physics, University of Oulu, P. O. Box 3000, FIN-90014 University of Oulu, Finland}
\address{$^f$ Universit\'e  Grenoble Alpes, CNRS, LIPhy, F-38000 Grenoble, France}
\address{$^g$ Institute  of Chemistry, E\"otv\"os Lor\'and University and MTA-ELTE Complex Chemical Systems Research Group,	H-1518 Budapest 112, P.O. Box 32, Hungary}

%\date{\today}	
\begin{abstract}
44~325 measured and assigned transitions of \hhs, the parent isotopologue of the hydrogen sulfide molecule, are collated from 33 publications into a single database and reviewed critically.
Based on this information, rotation-vibration energy levels are determined for the ground electronic state using the Measured Active Rotational-Vibrational Energy Levels (\Marvel) technique. 
The \textit{ortho} and \textit{para} principal components 
of the measured spectroscopic network of \hhs\ are considered separately. 
The verified set of 25~293 \textit{ortho-} and 18~778 \textit{para-} H$_2$$^{32}$S transitions
determine 3969 \textit{ortho} and 3467 \textit{para} energy levels. 
The \Marvel\ results are compared with alternative data compilations,
including a theoretical variational linelist.
\end{abstract}
%begin{keywords} Acetylene \sep MARVEL \sep rotional-vibration transitions \sep energy levels \end{keywords}
\maketitle

\newpage
\section{Introduction}\label{sec:intro}

Hydrogen sulfide, H$_2$S, is a volcanic gas present on earth and other 
geologically active planets and moons, such as Io \cite{90SaAlWi.H2S}, 
Venus \cite{04ScGrAb.H2S}, and, theoretically, hot super-earth "lava" exoplanets \cite{jt693}. It has recently been detected above the clouds in the atmosphere of Uranus~\cite{18IrToGa.H2S}.
On earth, microbial respiration of seawater sulfate creates a hydrogen sulfide-rich 
ecosystem \cite{97Pace.H2S}, suggesting that H$_2$S is a component for a potential 
microbial life sustaining atmosphere on Venus \cite{04ScGrAb.H2S},
and a potential biomarker for life on exoplanets \cite{10KaSaxx}.
H$_2$S influences many physiological processes \cite{13King.H2S}, 
it is of importance in the treatment of respiratory diseases \cite{17BaAnIv.H2S}, 
and it is used as a measure of the quality of air near oil refineries \cite{08CiCuDe.H2S}.
The parent isotopologue, \hhs, was first detected in interstellar space in 
1972 \cite{72ThKuPe.H2S}. Many of the scientific and engineering applications mentioned require the detailed knowledge of the rovibrational energy levels of \hhs.

As part of this work, we present the largest compilation
of published experimental rovibrational transition data for \hhs.
The experimental database of H$_2$$^{32}$S transitions has been formatted 
and analysed using the \Marvel\ (Measured Active Rotational-Vibrational Energy Levels)
spectroscopic network (SN) software \cite{jt412,12FuCsxx.marvel,12FuCsa}.
%the results of which are presented and discussed in this paper. 
This study builds on the 2012 W@DIS information system for the hydrogen sulfide 
molecule \cite{12PoLaVo.H2S}. 

As to the structure of this paper, 
the next section provides the theory underlying the present study. 
Section~\ref{sec:sources} presents and discusses the experimental sources used, with results given in Section~\ref{sec:results}. 
Section~\ref{sec:comp} discusses these results and compares the empirical rovibrational energies 
presented in this work with corresponding levels previously determined both by 
experiment and theory. 
Finally, Section~\ref{sec:conc} provides our conclusions. 
All transition data and energy levels resulting from this work are included 
as supplementary data files.

\section{Theory}

\subsection{\Marvel}
The \Marvel\ procedure \cite{jt412,12FuCsxx.marvel,12FuCsa} is  based on the theory of
spectroscopic networks \cite{11CsFuxx.marvel,16ArPeFu.marvel}, 
with the energy levels represented as nodes and the transitions between them as edges. 
The related \Marvel\  code can be used to critically evaluate and validate 
experimentally-determined transition wavenumbers and uncertainties collected 
from the literature, inverting the wavenumber information to obtain accurate 
empirical energy levels with an associated uncertainty. 
\Marvel\ has been successfully used to evaluate the energy levels for molecules 
such as $^{12}$C$_2$ \cite{jt637}, $^{48}$Ti$^{16}$O \cite{jt672}, 
water vapour \cite{jt454,jt482,jt539,jt576,jtwaterupdate}, 
H$_3^+$ \cite{13FuSzMa.marvel}, H$_2$D$^+$ and D$_2$H$^+$ \cite{13FuSzFa.marvel}, 
$^{12}$C$_2$H$_2$ \cite{jt705}, $^{14}$NH$_3$ \cite{jt608,jtNH3update}, and $^{12}$C$_2$H$_2$$^{16}$O \cite{11FaMaFu.marvel}. 
\Marvel\ requires each measured transition to have an associated uncertainty and 
for each energy level considered to possess a unique set of quantum numbers. 
%The quantum numbers used in the present study are described in the following subsection.

\subsection{Quantum number labelling}
\label{QN_label}

The seven quantum numbers that were used for labelling the upper and lower rovibrational states of \hhs\ are the same as those used in a previous \Marvel\ investigation to classify water \cite{jt454}.
%, also of the $C_{2\rm{v}}$(M) molecular symmetry (MS) group \cite{98BuJe.method}. 
These quantum numbers are summarised in Table \ref{t:QN}. Normal-mode labelling is used for the vibrations, where $\nu_1, \nu_2$ and $\nu_3$ stand for the symmetric stretch, bend, and antisymmetric stretch vibrations, respectively.
Standard asymmetric-top quantum numbers are used for the rotations, where $J, K_a$, and $K_c$ are the three quantum numbers associated with rotational angular momentum, ${\bf J}$, and the two projections along the $A$ and $C$ axes.
%, $K_a$ and $K_c$. 
We also provide, as part of the label, the nuclear spin state (\textit{ortho} or \textit{para}), which is deduced by whether ($v_3+K_a+K_c$) is odd (\textit{ortho}) or even (\textit{para}) \cite{jt705,82HeLixx.C2H2}. Hyperfine coupling associated with nuclear spins has been neglected.

%In the alternative local-mode labelling the vibrations are described by ($n,m$)$^{\pm}$$v_2$, 
%with $v_2$ the same bending quantum number as in the normal-mode representation. The stretching-mode quantum numbers are given by $n+m=v_1+v_3$, with the superscripted sign indicating whether this is a symmetric or antisymmetric combination (with only the symmetric combination allowed for $n=m$). 
%When $n=m$, only the symmetric combination is allowed and so the superscript notation is dropped. 
When local-mode notation was used by an experimental source considered in this work, the quantum numbers were transformed to the
normal-mode notation $(v_1v_2v_3)$ to describe a vibrational state.
Note that some of the data sources considered in this study use $J$ and $\tau=K_a-K_c$ instead of $J, K_a$, and $K_c$ \cite{56AlPl.H2S}, 
%In this notation, $\tau=K_a-K_c$ and $K_a+K_c=J, J+1$, 
where $\tau$ runs from $-J, -J+1, ..., J-1, +J$, with energy increasing from $\tau=-J$. 
%The lowest-energy state would have $K_a=0$ and $K_c=J$, the second lowest $K_a=1$ and $K_c=J$, and so on.
The $+/-$ parity of an asymmetric top molecule such as \hhs\ is defined by $(-1)^{K_c}$ \cite{jt454}. \hhs\ belongs to the $C_{2\rm{v}}$(M) MS group \cite{98BuJe.method}, which contains irreducible representations A$_1$, A$_2$, B$_1$, and B$_2$, as given in Table~\ref{t:parity}. 
%According to our labelling scheme, these correspond to the four pairs: \textit{para} $+$, \textit{para} $-$, \textit{ortho} $-$ and \textit{ortho} $+$, respectively, as given in Table~\ref{t:parity}.  

\begin{table}
\caption{Quantum numbers used to label the upper and lower energy states of \hhs.} 
\label{t:QN}
\begin{tabular}{ll}
\hline\hline
Label & Description\\
\hline
$v_1$	& S-H symmetric stretch ($\sim$ 2614.4\cm)\\
$v_2$	& Symmetric bending mode ($\sim$ 1182.6\cm)\\
$v_3$	& S-H antisymmetric stretch ($\sim$ 2628.5\cm) \\
$J$	& Rotational angular momentum \\
$K_a,K_c$	& Projections of rotational angular momentum \\
\textit{ortho}/\textit{para}	& Nuclear spin state (see text) \\
\hline\hline
\end{tabular}
\end{table}

\begin{table}[H]
	\centering
	\caption{Symmetry of the rovibrational states of \hhs}
	\label{t:parity}
	\begin{tabular}{lcccc}
		\hline\hline
		Symmetry & A$_1$ & A$_2$ & B$_1$ & B$_2$ \\
		\hline
		Parity & $+$ & $-$ & $-$ & $+$ \\
		Nuclear spin state & \textit{para} & \textit{para} & \textit{ortho} & \textit{ortho} \\
		\hline\hline
	\end{tabular}
\end{table}

\subsection{Selection rules}

The rigorous selection rules governing rotation-vibration transitions for a
molecule of the $C_{2\rm{v}}$(M) MS group are given by:
\begin{align}
\Delta J =0, \pm 1, \\
J'+J'' \ne  0, \\
- {\leftrightarrow} +
\end{align}

%These selection rules have been checked when assembling the \Marvel\ database of experimental transitions.
The \textit{ortho} states of \hhs\ have the nuclear spin statistical 
weight $g_{\rm ns}=3$, while for the \textit{para} states $g_{\rm ns}=1$, 
%$+/-$ refers to the parity, $(-1)^{K_c}$ for an asymmetric top molecule such as \hhs. 
%The $g_{\rm ns}$ factors enter the intensity formula of the rovibrational transitions; 
thus, \textit{ortho} transitions have three times the intensity of \textit{para} transitions.
This is sometimes referred to as intensity alternation. 
It is assumed that \textit{ortho} and \textit{para} states do not interconvert. 
Such transitions are very weakly allowed \cite{jt329} but have yet to be observed for \hhs.

\section{Experimental sources}
\label{sec:sources}

A large number of experimentally determined transition wavenumbers can be found 
in the literature for the main isotopologue of hydrogen sulfide, H$_2$$^{32}$S.  
We have attempted to conduct a rigorous and comprehensive search 
for all useable spectroscopic data. 
Fortunately, much of the data up to 2012 was previously collated 
as part of the 2012 W@DIS information system for hydrogen sulfide \cite{12PoLaVo.H2S}, 
in which some of the authors of this paper were involved. 
These data were converted to \Marvel\ format for this work and analysed alongside 
data from newly collected sources. 
This requires the transition wavenumber (in \cm) and the associated uncertainty, 
along with quantum number assignments for both the upper and lower energy states, 
and a unique reference label for each transition. 
This reference indicates the data source
%, table and line number that 
the transition originates from. 
The data source tag is based on the notation employed by an IUPAC Task Group 
on water spectroscopy \cite{jt454}. 
An extract of the input file in the required format is given in Table \ref{t:ortho}; 
the full file can be found in the supplementary data for this work. 

33 sources of experimental data were used in the final data set. The data from more recent papers are generally provided in digital format, but some of the older papers had to be processed through digitalisation software, or even manually entered in the worst 
cases. After digitalisation the data were converted to \Marvel\ format, as described above.

\begin{table}
\caption{Extract from the \Marvel\ input file for the \textit{ortho} transitions for \hhs. 
The full file is supplied as part of the supplementary information to this paper. 
All energy term values and uncertainties are in units of cm$^{-1}$. 
The meaning of the upper and lower state assignments can be found in Table \ref{t:QN}.
		\label{t:ortho}
	}\tt\footnotesize
	\begin{tabular}{rllllr}
		\hline\hline
		Transition & Uncertainty & Upper state assignment & Lower state assignment & Reference  \\
		\hline
33.12631	&	0.00015	&	0	0	0	5	2	3	ortho	&	0	0	0	5	1	4	ortho	&	94YaKl\_1	\\
33.12631	&	0.00018	&	0	0	0	3	0	3	ortho	&	0	0	0	2	1	2	ortho	&	94YaKl\_2	\\
34.00529	&	0.00010	&	0	0	0	8	7	2	ortho	&	0	0	0	8	6	3	ortho	&	94YaKl\_5	\\
34.15779	&	0.00010	&	0	0	0	11	9	2	ortho	&	0	0	0	11	8	3	ortho	&	94YaKl\_7	\\
34.21980	&	0.00010	&	0	0	0	6	4	3	ortho	&	0	0	0	6	3	4	ortho	&	94YaKl\_8	\\
34.24062	&	0.00012	&	0	0	0	4	2	3	ortho	&	0	0	0	4	1	4	ortho	&	94YaKl\_9	\\
35.73512	&	0.00010	&	0	0	0	8	6	3	ortho	&	0	0	0	8	5	4	ortho	&	94YaKl\_13	\\
		\hline
	\end{tabular}
\end{table}

Table \ref{t:sources} gives a summary of all the data sources used in this work, 
along with the energy range, number of transitions (assigned (A) and verified (V)), and comments, which are detailed in Section~\ref{sec:comments}.  
%The comments are detailed in Section~\ref{sec:comments}. For the temperatures, 
%RT stands for room temperature, and NS for ``not stated'' if the temperature is 
%not explicitly given in the original paper. 
%In these cases the equipment used is noted instead, with FT standing for Fourier Transform,
%VECSEL for `Vertical External Cavity Surface Emitting Laser', and 
%ICLAS for `Intracavity Laser Absorption Spectroscopy'. 
Table \ref{t:notused} lists those data sources which were considered but not used, 
with comments on the reasons for their exclusion from the analysis. 
The IUPAC reference tag given in these tables matches those used in the unique labels 
in the \Marvel\ input files, given in the supplementary data and illustrated 
in Table \ref{t:ortho}. 

As transitions have never been observed between \textit{ortho} and \textit{para} 
states of \hhs, they form two separate principal components (PCs) of 
the experimental spectroscopic network.
%, with no links between them. 
All input and output files supplied in the supplementary data to this work are 
split into either \textit{ortho} or \textit{para}.

{\footnotesize
\begin{longtable}{llcccl}
	\caption{Data sources used in this study with frequency range, numbers of transitions (A/V for assigned/verified), and comments, which are detailed in Section~\ref{sec:comments}. As far as we are aware all experiments were conducted at room temperature.
		\label{t:sources}} \\
	\hline
	\cr Tag &	Ref.	&	Range (cm$^{-1}$)	&	A/V  & Comments \\
	\hline\hline
	\endhead
72HeCoLu 	&	 \cite{72HeCoLu.H2S} 	&	 1.17--25.55 	&	 37/35 	 	&	  (4a)  \\   
95BeYaWiPo 	&	 \cite{95BeYaWi.H2S} 	&	 4.39--85.41 	&	 112/84 	  	&	  (4b) \\     
68CuKeGa 	&	 \cite{68CuKeGa.H2S} 	&	 5.63--14.15 	&	 6/6 	   	&	   \\    
71Huiszoon 	&	 \cite{71Huiszoon.H2S} 	&	 5.63--7.23 	&	 2/2 	   	&	   \\    
14CaPu 	&	 \cite{14CaPu.H2S} 	&	 7.23--53.17 	&	 70/70 	   	&	   \\   
85BuFeMeSh 	&	 \cite{85BuFeMe.H2S} 	&	 10.02--20.9 	&	 6/6 	  	&	   \\     
94YaKl 	&	 \cite{94YaKlxx.H2S} 	&	 33.13--259.76 	&	 366/366 	  	&	   \\    
13CaPu 	&	 \cite{13CaPu.H2S} 	&	 33.97--37.45 	&	 4/4 	 	&	    \\   
13AzYuTeMa 	&	 \cite{jt558} 	&	 45.25--359.79 	&	 1158/1139 	  	&	  (4c) \\   
83FlCaJo 	&	 \cite{83FlCaJo.H2S} 	&	 50.77--307.51 	&	 426/387 	  	&	  (4d) \\    
18UlBeGr 	&	 \cite{18UlBeGr.H2S} 	&	 729.78--1735.41 	&	 2267/2267 		&	   \\            
82LaEdGiBo 	&	 \cite{82LaEdGi.H2S} 	&	 1003.46--1495.28 	&	 397/396 	   	&	 (4e)  \\    
83Strow 	&	 \cite{83Strow.H2S} 	&	 1082.03--1257.07 	&	 123/123 	  	&	   \\    
96UlMaKoAl 	&	 \cite{96UlMaKo.H2S} 	&	 1178.05--1359.78 	&	 41/41 	  	&	   \\     
98BrCrCrNa 	&	 \cite{98BrCrCr.H2S} 	&	 2141.30--4249.85 	&	 7473/7473 	  	&	   \\     
18Horneman 	&	 \cite{17Horneman.H2S} 	&	 2180.35--4220.46 	&	4460/4460		&	   \\     
84LeFlCaJo 	&	 \cite{84LeFlCa.H2S} 	&	 2180.36--2945.81 	&	 2113/2111 	 	&	  (4f) \\     
81GiEd 	&	 \cite{81GiEd.H2S} 	&	 2192.48--2823.11 	&	 715/704 	&	  	  (4g) \\    
96UlOnKoAl 	&	 \cite{96UlOnKo.H2S} 	&	 3614.70--3887.66 	&	 106/106 	 	&	   \\     
05UlLiBeGr 	&	 \cite{05UlLiBe.H2S} 	&	 4000.59--6653.79 	&	2347/2347	   	&	   \\   
97BrCrCrNa 	&	 \cite{97BrCrCr.H2S} 	&	 4500.88--5595.01 	&	 5221/5219 	 	&	(4h)   \\ 
18Liu 	&	 \cite{18Liu.H2S} 	&	 4514.79--5555.58 	&	3337/3335		 	&	(4i) \\
04BrNaPoSi\_c 	&	 \cite{04BrNaPoa.H2S} 	&	 5688.27--6676.71 	&	3178/3178	 	&	 \\ 
04BrNaPoSi\_a 	&	 \cite{04BrNaPoSi.H2S} 	&	 7169.19--7898.97 	&	 2878/2876   	&	(4j)   \\
04UlLiBeGr\_b 	&	 \cite{04UlLiBe.H2S} 	&	 7226.81--7994.09 	&	1855/1855 	&	  \\   
04UlLiBeGr\_a 	&	 \cite{04UlLiBeb.H2S} 	&	 8405.91--8905.28 	&	 589/589  	&	   \\    
04BrNaPoSi\_b 	&	 \cite{04BrNaPob.H2S} 	&	 8412.73--8906.11 	&	1179/1175  	&	 (4k)  \\
03DiNaHuZh 	&	 \cite{03DiNaHu.H2S} 	&	 9541.01--10000.71 	&	 1736/1728 		&	  (4l) \\  
01NaCa\_a 	&	 \cite{01NaCaxxa.H2S} 	&	 10787.33--11297.99 	&	 1105/1097 	 	&	 (4m)  \\    
94GrRaStDe 	&	 \cite{94GrRaSt.H2S} 	&	 11948.91--12246.28 	&	 227/145 	 	&	   (4n) \\    
97VaBiCaFl 	&	 \cite{97VaBiCa.H2S} 	&	 12324.55--12670.68 	&	 399/387 	  	&	 (4o)  \\     
99CaFl 	&	 \cite{99CaFlxx.H2S} 	&	 13060.51--13357.14 	&	 219/206 	  	&	  (4p) \\   
01NaCa\_b 	&	 \cite{01NaCaxxb.H2S} 	&	 16186.25--16436.57 	&	 173/154 	  	&	  (4q) \\    
	\hline
Total &  & 1.17--16436.57 & 44~325/44~071 &       &   \\
	\hline
\end{longtable}
}

{\footnotesize
\begin{longtable}{lll}
	\caption{Data sources considered but not used in this work.
	\label{t:notused}} \\
	\hline
	\cr Tag	&	Ref.	&	Comments	\\
	\hline\hline
	\endhead
	02CoRoTy	&	\cite{02CoRoTy.H2S}	& Data taken from other sources	\\
	94WaKuSu	&	\cite{94WaKuSu.H2S}	& Data taken from other sources	\\
	96SuMeKr	&	\cite{96SuMeKr.H2S}	& Data taken from other sources	\\
	97SuMeKr	&	\cite{97SuMeKr.H2S}	& Data taken from other sources	\\
	97Sumpf	&	\cite{97Suxxxx.H2S}	& Data taken from other sources	\\
	98PiPoCoDe	&	\cite{98PiPoCo}	&  Data from http://spec.jpl.nasa.gov. Data does not appear to be experimental.\\
	73HeDeKi	&	\cite{73HeDeKi.H2S}	& Data taken from other sources, \cite{68CuKeGa.H2S,72HeCoLu.H2S,71Huiszoon.H2S} \\
	02KiSuKrTi	&	\cite{02KiSuKr.H2S}	& Data appears to be taken from the 1996 edition of HITRAN \cite{98RoRiGo.db}	\\
	06Polovtseva	&	N/A	&	Thesis, no citation. Referenced in \cite{12PoLaVo.H2S}	\\
	87LeFlCaAr	&	\cite{87LeFlCa.H2S}	&	Energy levels	\\
	13Azzam	&	\cite{13Azzamx.H2S}	& All results are given in 13AzYuTeMa \cite{jt558}\\
	94KoJe	&	\cite{94KoJexx.H2S}	&	Energy levels \\
	95FlGrRa	&	\cite{95FlGrRa.H2S}	&	Energy levels \\
	85LaEdGiBo    & \cite{85LaEdGi.H2S} & Calculated values, compared to experimental values from 83FlCaJo \cite{83FlCaJo.H2S}\\
	94ByNaSmSi    & \cite{94ByNaSm.H2S} & Energy Levels \\
	01TyTaSc &  \cite{01TyTaSc.H2S} & Vibrational energy Levels \\
	69MiLeHa & \cite{69MiLeHa.H2S} & Could not be validated with the other sources to a reasonable accuracy (see Section~\ref{sec:gen_comments}) \\    
	56AlPl & \cite{56AlPl.H2S} & Could not be validated with the other sources to a reasonable accuracy (see Section~\ref{sec:gen_comments}) \\   
	69SnEd & \cite{69SnEd.H2S} & Could not be validated with the other sources to a reasonable accuracy (see Section~\ref{sec:gen_comments}) \\  
	\hline
\end{longtable}
}

\subsection{Comments on the experimental sources of Table \ref{t:sources}}\label{sec:comments}

\noindent
\textbf{(4a)} 72HeCoLu \cite{72HeCoLu.H2S} contains 2 lines which have been cited as taken from other experimental sources, for which the original data are already in our dataset. 
These duplicates were removed. \\
\textbf{(4b)} 95BeYaWiPo \cite{95BeYaWi.H2S} contains 28 lines which have been cited as from other experimental sources, for which the original data are already in our dataset. These duplicates were removed. \\
\textbf{(4c)} 13AzYuTeMa \cite{jt558} gives data for rotational lines recorded 
at room temperature. 
The equipment used provides spectra with very high sensitivity and this 
enabled transitions between highly rotationally excited states to be recorded, 
even at room temperature. 
These high rotational excitations may not be so accurately 
assigned as those of lower energy, as there are no data from other sources in the same region to confirm these measurements. 
The supplementary data from the original paper does not contain the original experimental
transitions; this was provided by the corresponding author. 
There were 19 lines which could not be validated against other data and 
so were removed from our dataset (see Section~\ref{sec:gen_comments}).\\
\textbf{(4d)} 83FlCaJo \cite{83FlCaJo.H2S} contains 39 lines which have been cited as from other sources which are already present in our dataset. 
The duplicates were removed. \\
\textbf{(4e)} 82LaEdGiBo \cite{82LaEdGi.H2S} contains one line which could not be validated and so has been removed from our dataset.\\
\textbf{(4f)} 84LeFlCaJo \cite{84LeFlCa.H2S} contains 2 lines which could not be 
validated against other more recent data and were removed from our dataset. 
All data from this source were recalibrated; 
the calibration factor employed was 0.999~999~746 below 2450 \cm\
and 1.000~000~0714 above 2450 \cm. \\
\textbf{(4g)} 81GiEd \cite{81GiEd.H2S} contains 11 lines which could not be validated against
more recent data and thus were removed from our dataset. \\
\textbf{(4h)} 97BrCrCrNa \cite{97BrCrCr.H2S} contains 2 lines with no corresponding energy in the AYT2 linelist (see Section~\ref{sec:comp}) and so were removed from our dataset.   \\ 
\textbf{(4i)} 18Liu \cite{18Liu.H2S} contains 2 lines with no corresponding energy in the AYT2 linelist and so were removed from our dataset.   \\ 
\textbf{(4j)} 04BrNaPoSi\_a \cite{04BrNaPoSi.H2S} contains 2 lines with no corresponding energy in the AYT2 linelist and so were removed from our dataset.   \\ 
\textbf{(4k)} 04BrNaPoSi\_b \cite{04BrNaPob.H2S} contains 4 lines which could not be validated and were removed from our dataset.\\
\textbf{(4l)} 03DiNaHuZh \cite{03DiNaHu.H2S} contains one blended line which has been 
removed from our dataset. 
A further 7 lines were found to include a level 
%(($v_1,v_2,v_3,J,K_a,K_c$)=(1,4,1,7,4,4)) 
with no corresponding energy 
in the AYT2 linelist and so were removed from our dataset. \\
\textbf{(4m)} 01NaCa\_a \cite{01NaCaxxa.H2S} contains 
8 transitions 
%with upper vibrational level ($v_1,v_2,v_3$)=(1,3,2) and 3 transitions 
%with upper vibrational level ($v_1,v_2,v_3$)=(2,3,1),
which were found to have no corresponding levels in the AYT2 linelist 
and so were removed from our dataset. \\
%No other sources contained data from these vibrational levels, 
%so it is assumed that they were misassigned. \\
\textbf{(4n)} 94GrRaStDe \cite{94GrRaSt.H2S} tentatively assigned the vibrational band 
in their data as ($v_1,v_2,v_3$)=(2,2,2); 
however, in 95FlGrRaSt \cite{95FlGrRa.H2S}, a paper published later with the same authors, 
the revised assignment is ($v_1,v_2,v_3$)=(3,0,2). 
We adopt the latter vibrational label in our dataset. 
No other data sources probe this vibrational band. 
The frequency is only given to 3 decimal places, so the uncertainty was altered to match. 
The original dataset has 82 transitions either labelled with an asterisk to indicate lines 
from the H$_2$$^{34}$S isotopologue, or labelled with a dagger to indicate lines 
which have been perturbed due to those from H$_2$$^{34}$S~\cite{94GrRaSt.H2S}. 
We have commented these out by adding a minus sign to the wavenumber
%start of the line 
and added the label $\_pt$ to the end.\\
\textbf{(4o)} 97VaBiCaFl \cite{97VaBiCa.H2S} contains 4 lines 
which could not be validated and so were removed from our dataset. 
A further 8 lines were found to contain a level 
%(($v_1,v_2,v_3,J,K_a,K_c$)=(1,0,4,8,7,2) or ($v_1,v_2,v_3,J,K_a,K_c$)=(1,0,4,8,8,1)) 
with no corresponding energy in the 
AYT2 linelist and so were removed from our dataset.\\ 
\textbf{(4p)} The assignment for 99CaFl\_185\_na\_ct from 99CaFl \cite{99CaFlxx.H2S} 
leads to different spin states (\textit{ortho}/\textit{para}) for the upper and lower state,
which is forbidden. 
We have removed this transition from our dataset. 
5 other lines could not be validated and so were removed from our dataset. 
7 lines were found to have no variational counterparts in the AYT2 linelist and so were removed. \\
\textbf{(4q)} 01NaCa\_b	\cite{01NaCaxxb.H2S} contains one line which could not be validated 
and so was removed from our dataset. 
%A further 2 lines were found to have upper levels (($v_1,v_2,v_3,J,K_a,K_c$)=(4,0,3,6,6,0),(3,0,4,6,6,0)) with no corresponding energy in the AYT2 linelist (see Section 5) and so were removed from our dataset. \\
A further 18 lines were found to contain levels with no corresponding energies in the AYT2 linelist and so were removed from our dataset. \\

\subsection{General comments}\label{sec:gen_comments}

All transitions which were considered but not processed in the final dataset have a minus sign in front of the 
transition wavenumber (indicating that \Marvel\ will ignore it) in the input files provided as supplementary material and are labelled with a comment at the end of the reference (see the comments of Section~\ref{sec:comments}). 
We used a cut-off of 0.035 \cm\ as the largest acceptable uncertainty and removed any transitions with an uncertainty greater than this this, with a note in  Section~\ref{sec:comments} to indicate how many transitions from a particular data source could not be validated to within this accuracy. 

The publications 97BrCrCrNa~\cite{97BrCrCr.H2S}, 04BrNaPoSi\_c~\cite{04BrNaPoa.H2S}, 04BrNaPoSi\_a~\cite{04BrNaPoSi.H2S}, and 04BrNaPoSi\_b~\cite{04BrNaPob.H2S} include just a short description of the assignment and modelling of the Fourier Rransform (FT) spectra between 4500 and 8900\cm. Detailed information on the theoretical treatment of these spectra will be published
separately in due course. Some of these transitions, between 4400--8000~\cm, were previously reported in
the HITRAN-2012~\cite{jt557} and HITRAN-2016~\cite{jt691} databases, while the second decade region, 8400--8900\cm, is presented
for the first time. It is also worth mentioning that the \Marvel\ dataset includes a considerable number of transitions which include new, unpublished energy levels. 
As only the upper energy levels of the transitions were reported in 05UlLiBeGr~\cite{05UlLiBe.H2S}, 04UlLiBeGr\_b~\cite{04UlLiBe.H2S}, and 04UlLiBeGr\_a~\cite{04UlLiBeb.H2S}, the corresponding transition wavenumbers were recovered from A.-W. Liu (18Liu~\cite{18Liu.H2S}) using the energy levels obtained from 97BrCrCrNa~\cite{97BrCrCr.H2S}, 04BrNaPoSi\_c~\cite{04BrNaPoa.H2S}, 04BrNaPoSi\_a~\cite{04BrNaPoSi.H2S}, and 04BrNaPoSi\_b~\cite{04BrNaPob.H2S}. For this reason, the transitions in the \Marvel\ dataset which are referred to as from 05UlLiBeGr~\cite{05UlLiBe.H2S}, 04UlLiBeGr\_b~\cite{04UlLiBe.H2S}, and 04UlLiBeGr\_a~\cite{04UlLiBeb.H2S} contain a larger number
of assignments than reported in the original publications. The data of the first hexad region between 4500--5600\cm\ were totally assigned based
on the energy levels reported in 97BrCrCrNa~\cite{97BrCrCr.H2S}.
The highly accurate (an accuracy of 0.0005\cm) set of \hhs\ transitions recorded using a FT spectrometer between
2200--4250\cm\ was provided by V.-M. Horneman (18Horneman~\cite{17Horneman.H2S}) and assigned using the energy levels reported in 98BrCrCrNa~\cite{98BrCrCr.H2S}.

\subsection{Sources from the HITRAN database}

\hhs\ has been included in the HITRAN 
database \cite{92RoGaTi.db,01RoBaBe.db,jt350,jt453,jt557,jt691}
since 1991 \cite{92RoGaTi.db}. 
The following are sources for the line positions and energy levels of the \hhs\ data 
in the HITRAN database, up to the 2016 release: 96UlMaKoAl \cite{96UlMaKo.H2S}, 83FlCaJo \cite{83FlCaJo.H2S}, 98BrCrCrNa \cite{98BrCrCr.H2S}, 13AzYuTeMa \cite{jt558}, 94YaKl \cite{94YaKlxx.H2S}, 95BeYaWiPo \cite{95BeYaWi.H2S}, 05UlLiBeGr \cite{05UlLiBe.H2S}, 04UlLiBeGr \cite{04UlLiBe.H2S}, 03DiNaHuZh \cite{03DiNaHu.H2S}, 01NaCa\_b \cite{01NaCaxxb.H2S}, 84LeFlCaJo \cite{84LeFlCa.H2S}, 82LaEdGiBo \cite{82LaEdGi.H2S}, 85LaEdGiBo \cite{85LaEdGi.H2S}, 94ByNaSmSi \cite{94ByNaSm.H2S}. 
The variational ExoMol AYT2 linelist \cite{jt640} was used to assign transitions 
in the $\nu_2$ excited vibrational state in the 2012 release \cite{jt557}.

\section{Results}
\label{sec:results}

The \Marvel\ website (\url{http://kkrk.chem.elte.hu/marvelonline}) has a version of \Marvel\ which can be run online. 
The variable NQN (number of quantum numbers) is 7 in the case of hydrogen sulfide, as illustrated in Table \ref{t:ortho} which 
shows an extract of the \textit{ortho} input file to \Marvel. 

%All energies are measured with respect to the zero-point energy (ZPE). 
%This is the energy of the ground ro-vibrational state, which is given a relative energy of 0 and is included in the \textit{para} set of energy levels. 
\Marvel\ automatically assigns the lowest energy state in a particular component of the spectroscopic network to 0. The ground rovibrational state is included in the \textit{para} set of energy levels, however there needs to be a ``magic number'', corresponding to the energy of the lowest \textit{ortho} state, which is added to all the \Marvel\   \textit{ortho}-symmetry energies. Here, this was taken as the ground vibrational state $(v_1v_2v_3)=(000)$ 
with the lowest rotational energy (see Section~\ref{QN_label}), $J=1, K_a=0, K_c=1$, of 94KoJe \cite{94KoJexx.H2S}, 
who determined the value of 13.74631 \cm. The output for the \textit{ortho} energies in the supplementary data, and the extract in  Table \ref{t:MARVEL}, all have this magic number added for the principal spectroscopic network. 
%The main component of the spectroscopic network in the \textit{para} output does not require a magic number as it contains the ground rovibrational level, $(000)$ $J=0, K_a=0, K_c=0$. 
There are a small number (4 \textit{para}) of energy levels which are not joined to either of the principal components. If more experimental transitions became available in the future it would be possible to link these to the principal network.

\begin{table}
	\caption{Extract from the \Marvel\ output file for the \textit{ortho} energy levels of \hhs. 
	The full file is supplied as part of the supplementary information to this paper. 
	All energies and uncertainties are in units of cm$^{-1}$. 
	NumTrans gives the number of transitions which are linked to that particular energy level. Assignments are as given by Table \ref{t:QN}.}
	\tt \footnotesize
	\setlength\tabcolsep{0.9ex}
	\label{t:MARVEL}
	\begin{tabular}{lllllllllll}
		\hline\hline
		\multicolumn{7}{c}{Assignment} & Energy & Uncertainty & NumTrans & Sym \\
		\hline
0	&	0	&	0	&	1	&	0	&	1	&	ortho	&	13.746310	&	0.000001	&	177	&	B1	\\
0	&	0	&	0	&	1	&	1	&	0	&	ortho	&	19.375630	&	0.000001	&	153	&	B2	\\
0	&	0	&	0	&	2	&	1	&	2	&	ortho	&	38.297765	&	0.000002	&	271	&	B2	\\
0	&	0	&	0	&	2	&	2	&	1	&	ortho	&	55.161605	&	0.000002	&	265	&	B1	\\
0	&	0	&	0	&	3	&	0	&	3	&	ortho	&	71.424233	&	0.000002	&	282	&	B1	\\
0	&	0	&	0	&	3	&	1	&	2	&	ortho	&	95.056325	&	0.000001	&	344	&	B2	\\
0	&	0	&	0	&	3	&	2	&	1	&	ortho	&	107.368228	&	0.000001	&	331	&	B1	\\
0	&	0	&	0	&	4	&	1	&	4	&	ortho	&	114.177613	&	0.000001	&	282	&	B2	\\
0	&	0	&	0	&	3	&	3	&	0	&	ortho	&	117.392015	&	0.000001	&	265	&	B2	\\
0	&	0	&	0	&	4	&	2	&	3	&	ortho	&	148.418340	&	0.000001	&	379	&	B1	\\
		\hline
	\end{tabular}
\end{table}

We collated and considered a total of 44~325 transitions from 33
experimental sources (25~474 \textit{ortho} and 18~851 \textit{para}). 
Of those 254 were found to be inconsistent (could not be validated to within 0.035 \cm) with others and so removed 
from the final data set, leaving a total of 44~071 transitions 
used as input into \Marvel\ (25~293 \textit{ortho} and 18~778 \textit{para}). 

\begin{figure}[H]
	\includegraphics[width=\linewidth]{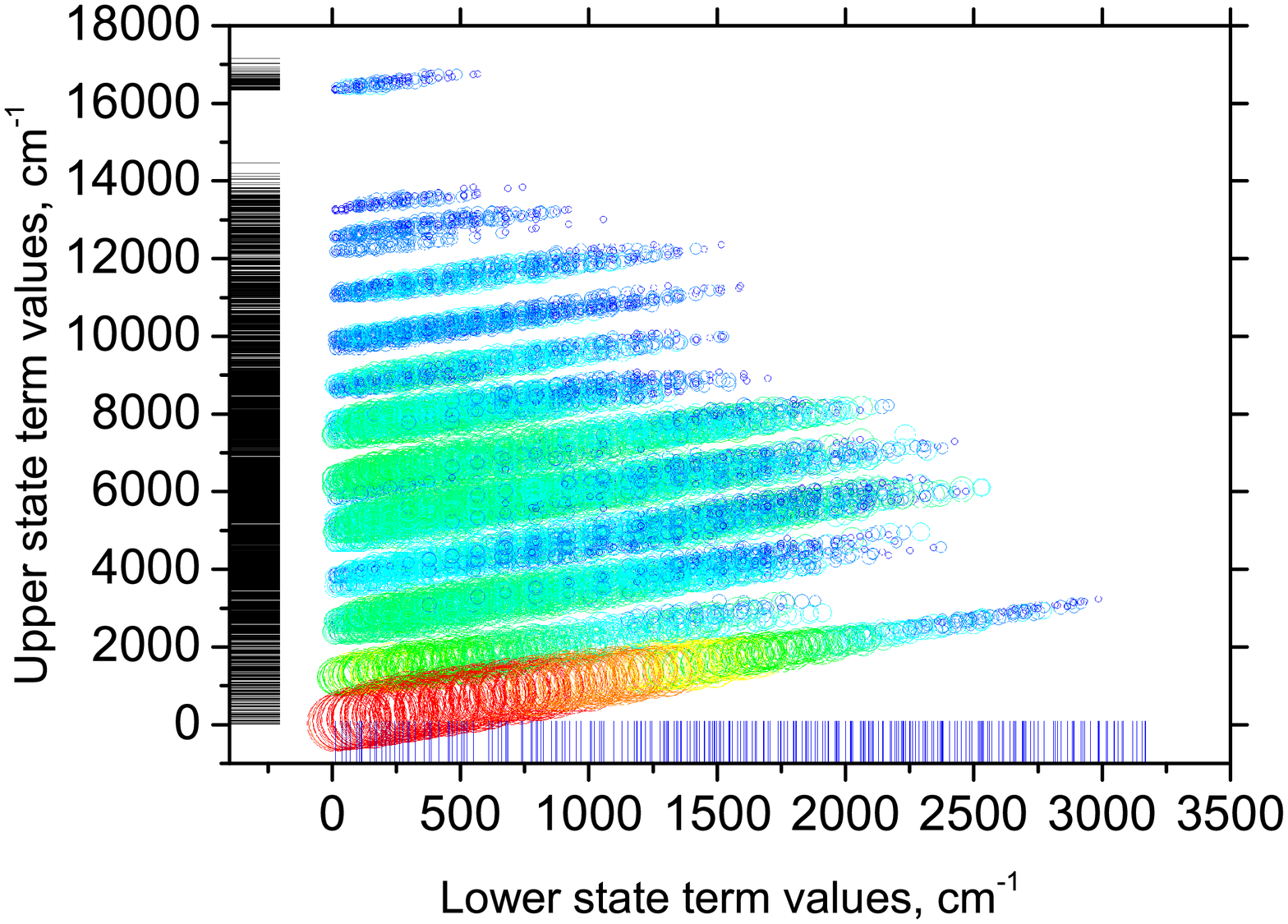}
	\caption{The lower-state energies of the experimental \textit{ortho}-\hhs\  transitions 
		used in this work, against corresponding upper-state energies. 
		The vertical bars along the horizontal-axis show the lower-state energies, 
		while the horizontal bars along the vertical-axis give the upper-state energies.  
		Each circle represents a particular transition, with the size proportional 
		to the log of NumTrans, the number of transitions supporting the upper state. 
		This value ranges from 1 (dark blue) to 404 (red).}
	\label{fig:H2S_CD}
\end{figure}

Figure \ref{fig:H2S_CD} gives a visual representation of NumTrans which link the states of the \Marvel\ \textit{ortho}-\hhs\ network, against corresponding upper state energies.
% The (blue) vertical bars along the horizontal-axis show the lower state energies, while the (black) horizontal bars along the vertical axis give the upper state energies. Each circle represents a particular transition, with the size proportional to the log of NumTrans.
%This value ranges from 1 (dark blue) to 405 (red). 
%As expected, the transitions between states of lower energies are supported by a higher number of transitions than those of higher energies. 
Those upper states which are dark blue in colour, linked by only 1 transition, should be considered less reliable than those in red, which are supported by hundreds of different transitions, up to 404. The values of NumTrans for each level are given in the energy level files in the supplementary data. 

%\begin{figure}[H]
%	\includegraphics[width=0.5\linewidth]{ortho_weighted_v7.png}
%	\includegraphics[width=0.5\linewidth]{para_weighted_v10.png}
%	\caption{The \textit{ortho} (left) and \textit{para} (right) components of the experimental spectroscopic network 
%		for \hhs\ produced using \Marvel\ input data}
%	\label{fig:ortho_spectro}
%\end{figure}

%These figures have been produced using algorithms from gephi, 
%an open source network visualisation software \cite{09MaHeJa}. 
%Different algorithms can be used to present these networks in a variety of ways. 
%Figure \ref{fig:orthopara_spectro}, for example, 
%gives alternative representations of the structure of the experimental spectroscopic network. 
%These figures highlight the intricate relationships between different energy levels 
%and illustrate how the variety of sources collated in this work link together. 
%If transition intensities were to be included as weights in the spectroscopic network 
%it could aid in the determination of transitions which should preferentially be 
%investigated in new experiments \cite{11CsFuxx.marvel}.

%
%\begin{figure}[H]
%\includegraphics[width=\linewidth]{ortho_weighted_v7.png}
%\caption{The \textit{ortho} component of the experimental spectroscopic network 
%for \hhs\ produced using \Marvel\ input data}
%\label{fig:ortho_spectro}
%\end{figure}
%
%\begin{figure}[H]
%\includegraphics[width=\linewidth]{para_weighted_v10.png}
%\caption{The \textit{para} component of the experimental spectroscopic network 
%for \hhs\  produced using \Marvel\  input data}
%\label{fig:para_spectro}
%\end{figure}

%\begin{figure}[H]
%	\includegraphics[width=0.5\textwidth]{ortho_v8.png}
%	\includegraphics[width=0.5\textwidth]{para_notweighted_1b.png}
%	\caption{Alternative representations of the \textit{ortho} (left) 
%	and \textit{para} (right) components of the spectroscopic network for \hhs\  produced 
%	using \Marvel\ input data.}
%	\label{fig:orthopara_spectro}
%\end{figure}

\section{Comparison to other derived energy levels}
\label{sec:comp}

We used a number of sources to compare energy level values 
against those determined in this work. 
94KoJe \cite{94KoJexx.H2S} contains a list of energy levels, which are taken 
or calculated from other 
sources \cite{83FlCaJo.H2S,85LaEdGi.H2S,84LeFlCa.H2S,69SnEd.H2S,56AlPl.H2S,87LeFlCa.H2S}; see Table \ref{t:96KoJe} for a breakdown by vibrational band. 
AYT2 \cite{jt640} is a theoretical variational linelist for \hhs\ calculated 
as part of the ExoMol project~\cite{jt528,jt631}, a database of theoretical 
linelists for molecules of astrophysical importance. 
AYT2 is appropriate up
to temperatures of around 2000 K and designed for use in characterising 
the atmospheres of cool stars and exoplanets. 
The states file, available from www.exomol.com, gives the calculated energy levels, 
which can be compared against our empirical energy levels. 
For more information on the format of the ExoMol data files, 
please refer to \cite{jt631}. 
95FlGrRaSt \cite{95FlGrRa.H2S} contains rovibrational energy levels from 
the ($v_1,v_2,v_3$)=(2,0,3) and (3,0,2) vibrational bands. 
The authors outline an experimental procedure used to deduce these levels 
but they do not provide the experimental transition data so we could not include 
the data from this source in our dataset. 
87LeFlCaAr \cite{87LeFlCa.H2S} has a table of rovibrational energy levels for 
the ($v_1,v_2,v_3$)=(2,1,0), (1,1,1) and (0,1,2) vibrational bands. 
96UlMaKoAl \cite{96UlMaKo.H2S} contains rovibrational energy levels from 
the ($v_1,v_2,v_3$)=(0,1,0) vibrational band.

{\footnotesize
	\begin{longtable}{llll}
		\caption{Original sources of energy level data from 94KoJe \cite{94KoJexx.H2S}.\label{t:96KoJe}} \\
		\hline\hline
		\cr Tag & Ref.	& Vibrational bands &	Comments	\\
		\hline
		83FlCaJo & \cite{83FlCaJo.H2S} & (0,0,0) & Experimental source used in this work\\
		85LaEdGiBo & \cite{85LaEdGi.H2S} & (0,1,0) & Calculated values, compared to experimental values from 83FlCaJo \cite{83FlCaJo.H2S} \\
		84LeFlCaJo & \cite{84LeFlCa.H2S} & (0,2,0) (1,0,0) (0,0,1) & Experimental source used in this work\\
		69SnEd & \cite{69SnEd.H2S} &  (1,1,0) (0,1,1) & Experimental source used in this work\\
		56AlPl & \cite{56AlPl.H2S} &  \specialcell[t]{(0,2,1) (2,0,0) (1,0,1) \\ (3,0,0) (2,0,1) (1,0,2) \\ (0,0,3) (2,1,1) (3,0,1) \\ (2,0,2) (1,0,3) (3,1,1)}      & Data source not considered reliable. All $J=0$ levels only.\\
		87LeFlCaAr & \cite{87LeFlCa.H2S} &  (2,1,0) (1,1,1) & \specialcell[t]{94KoJe \cite{94KoJexx.H2S} only contains levels from this source up to $J=5$.\\See text and Figure \ref{fig:energy_levels} for a full comparison.}\\
		\hline
	\end{longtable}
}

We compared the pure rotational levels determined in this work both against those 
given in 94KoJe \cite{94KoJexx.H2S} and the calculated values from the AYT2 
linelist \cite{jt640}, up to $J$=5. 
These data are given in Table \ref{t:compare}. 
The pure rotational levels are closer to those from the experimentally determined 
sources of 94KoJe than to those of variationally calculated AYT2, as would be expected. 

{\footnotesize
	\begin{longtable}{lcccrrrrrrr}
		\caption{Comparison of pure rotational levels from this work with those of 
		94KoJe \cite{94KoJexx.H2S} (see Table~\ref{t:96KoJe}) and AYT2 \cite{jt640} up to $J$=5. 
		NumTrans gives the number of transitions linking a particular state. 
		All energies and uncertainties are in \cm. 
		The uncertainty refers to this work.
			\label{t:compare}} \\
		\hline\hline
$J$  $K_a$  $K_c$ & State & Sym & NumTrans & Uncertainty & This work & 94KoJe & Difference & AYT2 & Difference \\
\hline
0 0 0  &  \textit{para}  &  A$_1$  & 58 & 0.000098 & 0 & 0 & 0 & 0 &  0 \\
1 0 1  &  \textit{ortho}  &  B$_1$  & 170 & 0.000001 & 13.746310 & 13.74631 & 0.00000 & 13.746340 &  0.000030 \\
1 1 1  &  \textit{para}  &  A$_2$  & 157 & 0.000002 & 15.090119 & 15.09011 & $-$0.00001 & 15.090151 &  0.000032 \\
1 1 0  &  \textit{ortho}  &  B$_2$  & 147 & 0.000001 & 19.375630 & 19.37563 & 0.00000 & 19.375563 &  $-$0.000067 \\     
2 0 2  &  \textit{para}  &  A$_1$  & 223 & 0.000001 & 38.016095 & 38.01607 & $-$0.00003 & 38.016264 &  0.000169 \\
2 1 2  &  \textit{ortho}  &  B$_2$  & 258 & 0.000002 & 38.297765 & 38.29775 & $-$0.00001 & 38.297961 &  0.000196 \\     
2 1 1  &  \textit{para}  &  A$_2$  & 211 & 0.000001 & 51.140188 & 51.14016 & $-$0.00003 & 51.140060 &  $-$0.000128 \\   
2 2 1  &  \textit{ortho}  &  B$_1$  & 253 & 0.000002 & 55.161605 & 55.16158 & $-$0.00003 & 55.161460 &  $-$0.000145 \\  
2 2 0  &  \textit{para}  &  A$_1$  & 181 & 0.000001 & 58.368870 & 58.36884 & $-$0.00003 & 58.368621 &  $-$0.000249 \\   
3 0 3  &  \textit{ortho}  &  B$_1$  & 265 & 0.000002 & 71.424221 & 71.42426 & 0.00004 & 71.424760 &  0.000539 \\
3 1 3  &  \textit{para}  &  A$_2$  & 213 & 0.000001 & 71.465192 & 71.46515 & $-$0.00004 & 71.465653 &  0.000461 \\
3 1 2  &  \textit{ortho}  &  B$_2$  & 329 & 0.000001 & 95.056313 & 95.05630 & $-$0.00001 & 95.056267 &  $-$0.000046 \\  
3 2 2  &  \textit{para}  &  A$_1$  & 272 & 0.000001 & 96.392529 & 96.39246 & $-$0.00007 & 96.392428 &  $-$0.000101 \\   
3 2 1  &  \textit{ortho}  &  B$_1$  & 313 & 0.000001 & 107.368216 & 107.36820 & $-$0.00002 & 107.367927 &  $-$0.000289 \\
4 0 4  &  \textit{para}  &  A$_1$  & 205 & 0.000001 & 114.172246 & 114.17217 & $-$0.00008 & 114.173085 &  0.000839 \\   
4 1 4  &  \textit{ortho}  &  B$_2$  & 270 & 0.000001 & 114.177601 & 114.17758 & $-$0.00002 & 114.178495 &  0.000894 \\  
3 3 1  &  \textit{para}  &  A$_2$  & 250 & 0.000002 & 115.340656 & 115.34059 & $-$0.00007 & 115.340038 &  $-$0.000618 \\
3 3 0  &  \textit{ortho}  &  B$_2$  & 247 & 0.000001 & 117.392003 & 117.39199 & $-$0.00001 & 117.391407 &  $-$0.000596 \\
4 1 3  &  \textit{para}  &  A$_2$  & 306 & 0.000001 & 148.140653 & 148.14058 & $-$0.00007 & 148.140740 &  0.000087 \\   
4 2 3  &  \textit{ortho}  &  B$_1$  & 362 & 0.000001 & 148.418328 & 148.41831 & $-$0.00002 & 148.418470 &  0.000142 \\  
5 0 5  &  \textit{ortho}  &  B$_1$  & 264 & 0.000045 & 166.343488 & 166.34345 & $-$0.00004 & 166.344887 &  0.001399 \\  
5 1 5  &  \textit{para}  &  A$_2$  & 196 & 0.000077 & 166.344005 & 166.34417 & 0.00016 & 166.345606 &  0.001601 \\      
4 2 2  &  \textit{para}  &  A$_1$  & 308 & 0.000001 & 170.335799 & 170.33574 & $-$0.00006 & 170.335471 &  $-$0.000328 \\
4 3 2  &  \textit{ortho}  &  B$_2$  & 383 & 0.000002 & 173.967266 & 173.96726 & $-$0.00001 & 173.966794 &  $-$0.000472 \\
4 3 1  &  \textit{para}  &  A$_2$  & 261 & 0.000001 & 182.648548 & 182.64849 & $-$0.00006 & 182.647978 &  $-$0.000570 \\
4 4 1  &  \textit{ortho}  &  B$_1$  & 290 & 0.000002 & 195.661414 & 195.66142 & 0.00001 & 195.659965 &  $-$0.001449 \\  
4 4 0  &  \textit{para}  &  A$_1$  & 219 & 0.000003 & 196.802189 & 196.80212 & $-$0.00007 & 196.800713 &  $-$0.001476 \\
5 1 4  &  \textit{ortho}  &  B$_2$  & 356 & 0.000037 & 210.217262 & 210.21727 & 0.00001 & 210.217731 &  0.000469 \\     
5 2 4  &  \textit{para}  &  A$_1$  & 270 & 0.000001 & 210.264821 & 210.26477 & $-$0.00005 & 210.265229 &  0.000408 \\   
5 2 3  &  \textit{ortho}  &  B$_1$  & 387 & 0.000001 & 243.343434 & 243.34344 & 0.00001 & 243.343211 &  $-$0.000223 \\  
5 3 3  &  \textit{para}  &  A$_2$  & 333 & 0.000001 & 244.392517 & 244.39250 & $-$0.00002 & 244.392149 &  $-$0.000368 \\
5 3 2  &  \textit{ortho}  &  B$_2$  & 363 & 0.000001 & 263.738931 & 263.73895 & 0.00002 & 263.738560 &  $-$0.000371 \\  
5 4 2  &  \textit{para}  &  A$_1$  & 298 & 0.000001 & 271.106067 & 271.10604 & $-$0.00003 & 271.104809 &  $-$0.001258 \\
5 4 1  &  \textit{ortho}  &  B$_1$  & 316 & 0.000002 & 277.337562 & 277.33758 & 0.00002 & 277.336633 &  $-$0.000929 \\  
5 5 1  &  \textit{para}  &  A$_2$  & 229 & 0.000002 & 296.104442 & 296.10442 & $-$0.00002 & 296.101282 &  $-$0.003160 \\
5 5 0  &  \textit{ortho}  &  B$_2$  & 278 & 0.000001 & 296.677577 & 296.67760 & 0.00002 & 296.674560 &  $-$0.003017 \\  
		\hline\hline
	\end{longtable}
}

The pure vibrational levels from this work are also compared with those from 
94KoJe \cite{94KoJexx.H2S} and AYT2 \cite{jt640}, see Table \ref{t:compare_vib}. 
It should be noted here that there are some differences in labelling between 
these data sources and those used in this work. 
For example, the pure vibrational levels of 94KoJe \cite{94KoJexx.H2S} which 
are labelled ($v_1,v_2,v_3$)=(2,0,2), (1,0,2) and (3,0,0), match the normal 
mode labelling of (4,0,0), (3,0,0) and (1,0,2), respectively, 
according to the labelling of 01TyTaSc \cite{01TyTaSc.H2S} and the experimental sources 
used in this work.

{\footnotesize
	\centering
	\begin{longtable}{cllrrrlrrr}
		\caption{Comparison of pure vibrational levels ($J=0$) from this work with those 
		of 94KoJe \cite{94KoJexx.H2S} (see Table~\ref{t:96KoJe}) and AYT2 \cite{jt640}. 
		NT stands for NumTrans, the number of transitions linking a particular state. 
		All energies and uncertainties are in \cm. 
		The uncertainty refers to this work. 
		State is from this work and Sym is the corresponding symmetry label as used 
		in AYT2 \cite{jt640} (see Section~\ref{QN_label}). 
			\label{t:compare_vib}} \\
		\hline\hline
$\nu_1$  $\nu_2$  $\nu_3$ & State & Sym & NT & Uncertainty & This work & 94KoJe & Difference & AYT2 & Difference \\
\hline
0 0 0  &  \textit{para}  &  A$_1$  & 58 & 0.000098 & 0 &    &    & 0 &  0 \\
0 1 0  &  \textit{para}  &  A$_1$  & 5 & 0.000477 & 1182.576991 & 1182.5742 &  $-$0.0028  & 1182.569618 &  $-$0.007373 \\
0 2 0  &  \textit{para}  &  A$_1$  & 4 & 0.000629 & 2353.964679 & 2353.9655 & 0.0008 & 2353.907317 &  $-$0.057362 \\    
1 0 0  &  \textit{para}  &  A$_1$  & 3 & 0.000663 & 2614.407743 & 2614.4074 &  $-$0.0003  & 2614.394829 &  $-$0.012914 \\
0 0 1  &  \textit{ortho}  &  B$_2$  & 2 & 0.000669 & 2628.454821 & 2628.4552 & 0.0004 & 2628.463320 &  0.008499 \\
0 3 0  &  \textit{para}  &  A$_1$  & 3 & 0.000621 & 3513.789974 &    &    & 3513.705072 &  $-$0.084902 \\
1 1 0  &  \textit{para}  &  A$_1$  & 3 & 0.000390 & 3779.166566 & 3779.1710 & 0.0044 & 3779.189348 &  0.022782 \\       
0 1 1  &  \textit{ortho}  &  B$_2$  & 2 & 0.000625 & 3789.269211 & 3789.2720 & 0.0028 & 3789.269878 &  0.000667 \\
0 4 0  &  \textit{para}  &  A$_1$  & 1 & 0.001000 & 4661.672219 &    &    & 4661.605794 &  $-$0.066425 \\
1 2 0  &  \textit{para}  &  A$_1$  & 2 & 0.000707 & 4932.699369 &    &    & 4932.688937 &  $-$0.010432 \\
0 2 1  &  \textit{ortho}  &  B$_2$  & 2 & 0.000707 & 4939.104010 & 4939.2300 & 0.1260 & 4939.129851 &  0.025841 \\
2 0 0  &  \textit{para}  &  A$_1$  & 2 & 0.000707 & 5144.986319 & 5145.1200 & 0.1337 & 5145.031868 &  0.045549 \\       
1 0 1  &  \textit{ortho}  &  B$_2$  & 2 & 0.000707 & 5147.220560 & 5147.3600 & 0.1394 & 5147.166622 &  $-$0.053938 \\   
0 0 2  &  \textit{para}  &  A$_1$  & 2 & 0.000707 & 5243.101919 &    &    & 5243.158956 &  0.057037 \\
1 3 0  &  \textit{para}  &  A$_1$  & 2 & 0.000928 & 6074.581067 &    &    & 6074.566059 &  $-$0.015008 \\
0 3 1  &  \textit{ortho}  &  B$_2$  & 2 & 0.000707 & 6077.594560 &    &    & 6077.626636 &  0.032076 \\ 
2 1 0  &  \textit{para}  &  A$_1$  & 2 & 0.000707 & 6288.146119 & 6288.1428 &  $-$0.0033  & 6288.134723 &  $-$0.011396 \\
1 1 1  &  \textit{ortho}  &  B$_2$  & 3 & 0.000637 & 6289.172875 & 6289.1739 & 0.0010 & 6289.128284 &  $-$0.044591 \\   
1 2 1  &  \textit{ortho}  &  B$_2$  & 4 & 0.000632 & 7420.092707 &    &    & 7420.077786 &  $-$0.014921 \\
1 0 2  &  \textit{para}  &  A$_1$  & 1 & 0.002000 & 7576.381719 & 7576.3000 &  $-$0.0817  & 7576.413281 &  0.031562 \\  
2 0 1  &  \textit{ortho}  &  B$_2$  & 2 & 0.001414 & 7576.544710 & 7576.3000 &  $-$0.2447  & 7576.596211 &  0.051501 \\ 
3 0 0  &  \textit{para}  &  A$_1$  & 2 & 0.001414 & 7752.263219 & 7751.9000 &  $-$0.3632  & 7752.343205 &  0.079986 \\  
0 0 3  &  \textit{ortho}  &  B$_2$  & 2 & 0.001414 & 7779.321260 & 7779.2000 &  $-$0.1213  & 7779.352004 &  0.030744 \\ 
1 3 1  &  \textit{ortho}  &  B$_2$  & 1 & 0.002000 & 8539.561310 &    &    & 8539.565999 &  0.004689 \\ 
1 1 2  &  \textit{para}  &  A$_1$  & 2 & 0.001414 & 8697.141469 &    &    & 8697.133905 &  $-$0.007564 \\
2 1 1  &  \textit{ortho}  &  B$_2$  & 2 & 0.001481 & 8697.154984 & 8697.3000 & 0.1450 & 8697.179341 &  0.024357 \\
1 4 1  &  \textit{ortho}  &  B$_2$  & 1 & 0.005000 & 9647.167310 &    &    & 9647.098855 &  $-$0.068455 \\
2 2 1  &  \textit{ortho}  &  B$_2$  & 1 & 0.005000 & 9806.667310 &    &    & 9806.712978 &  0.045668 \\ 
1 2 2  &  \textit{para}  &  A$_1$  & 1 & 0.005000 & 9806.733119 &    &    & 9806.748170 &  0.015051 \\
2 0 2  &  \textit{para}  &  B$_2$  & 1 & 0.005000 & 9911.023119 & 9911.0500 & 0.0269 & 9911.102285 &  0.079166 \\       
3 0 1  &  \textit{ortho}  &  A$_1$  & 1 & 0.005000 & 9911.023310 & 9911.0500 & 0.0267 & 9911.112478 &  0.089168 \\
3 1 1  &  \textit{ortho}  &  A$_1$  & 1 & 0.005000 & 11008.695310 & 11008.7800 & 0.0847 & 11008.774494 &  0.079184 \\   
3 0 2  &  \textit{para}  &  A$_1$  & 1 & 0.001000 & 12149.460119 &    &    & 12149.552318 &  0.092199 \\
1 0 4  &  \textit{para}  &  A$_1$  & 1 & 0.015000 & 12524.637119 &    &    & 12524.834491 &  0.197372 \\
4 0 1  &  \textit{ortho}  &  B$_2$  & 1 & 0.015000 & 12525.214310 &    &    & 12525.346292 &  0.131982 \\
3 1 2  &  \textit{para}  &  B$_2$  & 1 & 0.015000 & 13222.762119 &    &    & 13222.790827 &  0.028708 \\
2 1 3  &  \textit{ortho}  &  A$_1$  & 1 & 0.015000 & 13222.772310 &    &    & 13222.798559 &  0.026249 \\
		\hline\hline
	\end{longtable}
}

\begin{figure}[H]
	\includegraphics[width=\linewidth]{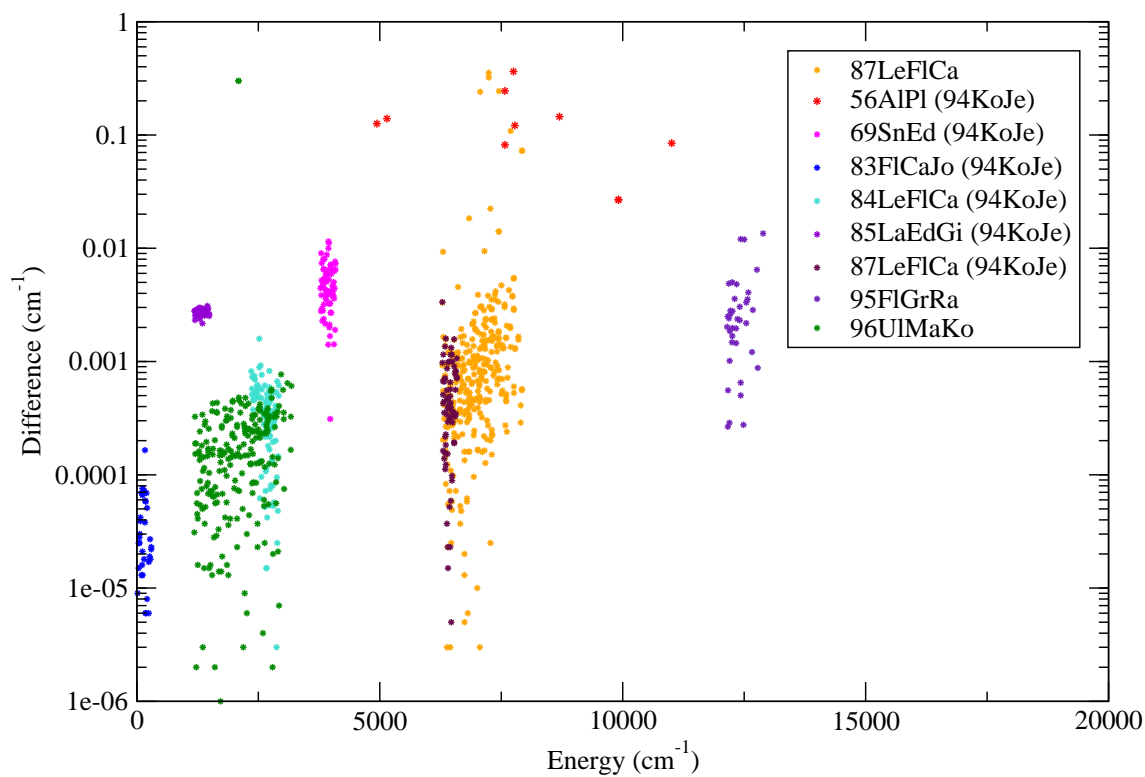}
	\caption{Deviations in cm$^{-1}$ between energy levels from this work and: 94KoJe \cite{94KoJexx.H2S} (see Table~\ref{t:96KoJe}), 95FlGrRaSt \cite{95FlGrRa.H2S}, 87LeFlCaAr \cite{87LeFlCa.H2S}, 96UlMaKoAl \cite{96UlMaKo.H2S}. Note the logarithmic vertical--axis.}
	\label{fig:energy_levels}
\end{figure}

We compared all the rovibrational energy levels given in 94KoJe \cite{94KoJexx.H2S}, 
95FlGrRaSt \cite{95FlGrRa.H2S}, 96UlMaKoAl \cite{96UlMaKo.H2S}, 
and 87LeFlCaAr \cite{87LeFlCa.H2S} against our set of \Marvel\ energy levels, 
taking the aforementioned labelling differences into account for
 94KoJe \cite{94KoJexx.H2S} (see Table \ref{t:96KoJe}). 
These comparisons are illustrated in Figure \ref{fig:energy_levels}. 
Most levels are in good agreement, with only a few levels with differences between 0.1--0.4~\cm. These levels with the largest differences are from 87LeFlCaAr \cite{87LeFlCa.H2S} and 56AlPl \cite{56AlPl.H2S}. 
All the levels which are compared to those of 87LeFlCaAr \cite{87LeFlCa.H2S} have several transitions which link them to other energy levels in our \Marvel\ dataset, and originate from different sources, which increases their reliability. This indicates that the lack of accuracy for these particular energies originates from 87LeFlCaAr \cite{87LeFlCa.H2S} and not the \Marvel\ dataset. The transition data from 56AlPl \cite{56AlPl.H2S} was considered for use in the current study, but the data could not be validated with that from newer sources (see Table~\ref{t:notused}), and so the energy levels are also assumed to be unreliable. 

%but there are a handful of outliers 
%of up to 8~\cm\ difference from 87LeFlCaAr \cite{87LeFlCa.H2S}. 
%All these levels have several transitions which link them to other 
%energy levels in our \Marvel\ dataset, and originate from different sources, 
%which increases their reliability. 
%This indicates that the lack of accuracy for these particular energies originates 
%from 87LeFlCaAr \cite{87LeFlCa.H2S} and not the \Marvel\ dataset. 
%The table of energy levels in 87LeFlCaAr \cite{87LeFlCa.H2S} is also not of the
%highest digital quality, so some numbers are ambiguous.

As mentioned above, AYT2 \cite{jt640} is a theoretical variational linelist 
for \hhs\ calculated as part of the ExoMol project~\cite{jt528,jt631}. 
Highly accurate experimental energy levels provide essential input for testing 
and improving theoretically calculated line positions such as those within AYT2. 
The ExoMol data format convention is to have a `states' and a `trans' file. 
The states file gives the calculated rovibrational energy levels, 
which can be compared against our empirical energy levels, 
and the trans file contains the Einstein--$A$ coefficients for transitions between 
these states \cite{jt631}. 
This format allows calculated linelists to be retrospectively updated using 
more reliable experimental energy levels in order to improve their accuracy, 
see \cite{jt570} for an example.  

\begin{figure}[H]
	\includegraphics[width=\linewidth]{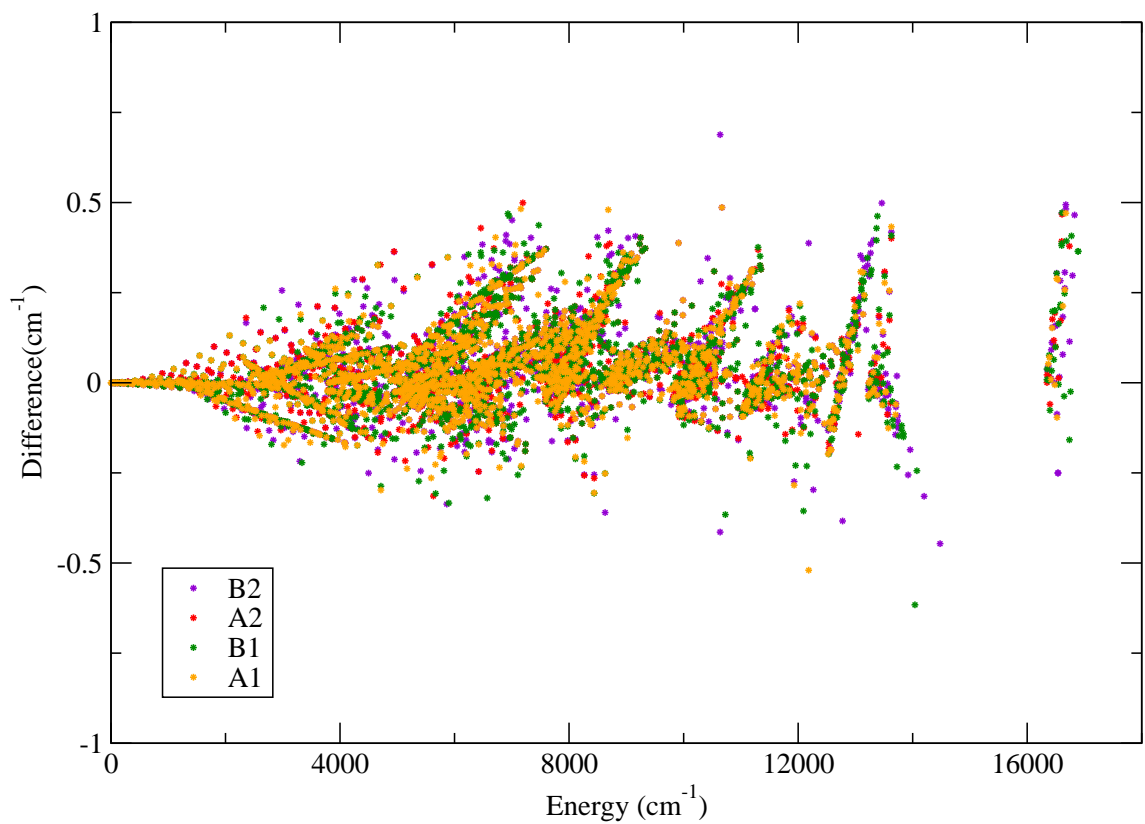}
\caption{Deviations, in cm$^{-1}$, between the MARVEL energy levels from this work 
and the variationally calculated linelist AYT2 \cite{jt640}. 
Different colors represent different rovibrational symmetries (see Table \ref{t:parity} 
in Section~\ref{QN_label}).}
\label{fig:exomol_levels}
\end{figure}

To compare our energy levels with those from the 
AYT2 \cite{jt640} linelist, a comparable states file was made; we labelled our states with the same A$_1$, A$_2$, B$_1$, B$_2$ symmetry 
labels which are given in the AYT2 states file, see Table \ref{t:parity}. 
We then matched all states with this same symmetry, 
$v_2$, $v_1+v_3$, and $J$, and searched for the closest value within 
these given parameters. 
Figure \ref{fig:exomol_levels} gives the result of this comparison. 
%We would generally be wary of levels with more than 1\cm\ difference in comparison with those from variational calculations. 

Work is underway to update the existing ExoMol AYT2 states file and linelist 
based on the energy levels obtained in this work, 
using only those levels based on the transitions we are most sure about. 
Care should be taken in general when using \Marvel\ energy levels 
with a low value of NumTrans
%, which stands for the number of transitions which are linked to that particular energy level 
(see Figure \ref{fig:H2S_CD}). 
This, along with the uncertainty, gives an indication of the reliability of a 
particular energy level.

\section{Conclusions}
\label{sec:conc}

A total of 44~325 measured experimental rovibrational transitions of \hhs\
from 31 publications have been considered in this work. 
From this set, 3969 \textit{ortho}- and 3467 \textit{para}-H$_2$$^{32}$S energy levels
have been determined using the Measured Active Rotational-Vibrational Energy
Levels (\Marvel) technique.
These results have been carefully compared with alternative compilations of energy levels.

A variational high-temperature linelist for \hhs\ has been computed as
part of the ExoMol project, AYT2 \cite{jt640}. 
Our new \Marvel\ energy levels will be used
to improve the accuracy of this theoretical linelist.

A significant part of this work was performed by  pupils from Highams Park
School in London, as part of a project known as ORBYTS (Original Research By
Young Twinkle Students) \cite{17McChRi}. 
The \Marvel\ studies of $^{12}$C$_2$H$_2$ \cite{jt705} and $^{48}$Ti$^{16}$O \cite{jt672} 
were also performed as part of the ORBYTS project
and further studies on other key molecules will be published in due course,
involving high-school students from the UK, Australia, and Hungary.
A paper discussing our experiences of performing original research in
collaboration with school children can be found elsewhere \cite{jt709}.

\section{Supplementary Data}\label{sec:fin}

Please refer to the web version of this work for access to the supplementary
data. There are four files provided, as listed in Table~\ref{t:SI}. The column definitions are given in Table~\ref{t:def} for files 1 and 2 (\Marvel\ input files) and Table~\ref{t:def} for files 3 and 4 (\Marvel\ output files).

\begin{table}[H]
	\caption{Supplied supplementary data files.}
	\label{t:SI}
	\tt
	\begin{tabular}{ll}
		\hline\hline
		File & Name\\
		\hline
		1 & H2S\_ortho\_input\_transitions.txt \\ 
		2 & H2S\_para\_input\_transitions.txt\\
		3 & H2S\_ortho\_energylevels\_output.txt\\ 
		4 & H2S\_para\_energylevels\_output.txt \\
		\hline\hline
	\end{tabular}
\end{table}

\begin{table}[H]
	\caption{Definition of columns in files 1 and 2.}
	\label{t:def}
	\footnotesize
	\begin{tabular}{lll}
		\hline\hline
		Column & Label & Description\\
		\hline
		1 & Energy (cm$^{-1}$)	& Transition wavenumber \\
		2 & Uncertainty (cm$^{-1}$)	& Associated uncertainity \\
		& Upper assignments: & \\
		3 & $v_1$	& S-H symmetric stretch \\
		4 & $v_2$	& Symmetric bending mode \\
		5 & $v_3$	& S-H antisymmetric stretch \\
		6 & $J$	& Rotational angular momentum \\
		7 & $K_a$	& Projection of rotational angular momentum \\
		8 & $K_c$	& Projection of rotational angular momentum \\
		9 & \textit{ortho}/\textit{para}	& Nuclear spin state \\
		& Lower assignments: & \\
		10 & $v_1$	& S-H symmetric stretch \\
		11 & $v_2$	& Symmetric bending mode \\
		12 & $v_3$	& S-H antisymmetric stretch \\
		13 & $J$	& Rotational angular momentum \\
		14 & $K_a$	& Projection of rotational angular momentum \\
		15 & $K_c$	& Projection of rotational angular momentum \\	
		16 & \textit{ortho}/\textit{para}	& Nuclear spin state \\
		17 & Ref & Unique reference label \\
		\hline\hline
	\end{tabular}
\end{table}

\begin{table}[H]
	\caption{Definition of columns in files 3 and 4.}
	\label{t:def34}
	\footnotesize
	\begin{tabular}{lll}
		\hline\hline
		Column & Label & Description\\
		\hline
		1 & $v_1$	& S-H symmetric stretch \\
		2 & $v_2$	& Symmetric bending mode \\
		3 & $v_3$	& S-H antisymmetric stretch \\
		4 & $J$	& Rotational angular momentum \\
		5 & $K_a$	& Projection of rotational angular momentum \\
		6 & $K_c$	& Projection of rotational angular momentum \\
		7 & \textit{ortho}/\textit{para}	& Nuclear spin state \\
		8 & Energy (cm$^{-1}$)	& \Marvel\ energy assignment \\
		9 & Uncertainty (cm$^{-1}$)	& \Marvel\ uncertainty \\
		10 & Num Trans	& The number of transitions in the dataset which link to this state \\
		\hline\hline
	\end{tabular}
\end{table}

\section{Acknowledgements}

We would like to thank Jon Barker, Fawad Sheikh, and Sheila Smith from Highams Park School 
for continued support and enthusiasm, along with Highams Park students Jack Franklin 
and Samuel Sheppard for their assistance in this project. 
We are grateful to Laura McKemmish for helpful discussions and advice, 
Clara Sousa-Silva for setting up the ORBYTS education project, 
and to Anita Heward, Will Dunn, Marcell Tessenyi, and the rest of the Twinkle team 
for their ongoing support. We give thanks to Shuiming Hu for contributions to the data used in this work. 
This work was supported by STFC Project ST/J002925, the RSF (grant 17-12-01204), 
the ERC under Advanced Investigator Project 267219, NKFIH (grant K119658), 
and the COST action MOLIM: Molecules in Motion (CM1405).

%%
%\bibliographystyle{elsarticle-num}
%\bibliography{journals_phys,jtj,C2H2,MARVEL,HCNO,linear,exogen,dwarfs,exoplanets,methods,H2S,linelists,ORBYTS} 

\begin{thebibliography}{10}
	\expandafter\ifx\csname url\endcsname\relax
	\def\url#1{\texttt{#1}}\fi
	\expandafter\ifx\csname urlprefix\endcsname\relax\def\urlprefix{URL }\fi
	\expandafter\ifx\csname href\endcsname\relax
	\def\href#1#2{#2} \def\path#1{#1}\fi
	
	\bibitem{90SaAlWi.H2S}
	F.~Salama, L.~Allamandola, F.~Witteborn, D.~Cruikshank, S.~Sandford,
	J.~Bregman, The 2.5-5.0 $\mu$m spectra of {Io}: Evidence for {H$_2$S} and
	{H$_2$O} frozen in {SO$_2$}, Icarus 83 (1990) 66 -- 82.
	\newblock \href {http://dx.doi.org/10.1016/0019-1035(90)90006-U}
	{\path{doi:10.1016/0019-1035(90)90006-U}}.
	
	\bibitem{04ScGrAb.H2S}
	D.~Schulze-Makuch, D.~H. Grinspoon, O.~Abbas, L.~N. Irwin, M.~A. Bullock, A
	sulfur-based survival strategy for putative phototrophic life in the
	{Venusian} atmosphere, Astrobiology 4 (2004) 11--18.
	\newblock \href {http://dx.doi.org/10.1089/153110704773600203}
	{\path{doi:10.1089/153110704773600203}}.
	
	\bibitem{jt693}
	J.~Tennyson, S.~N. Yurchenko, Laboratory spectra of hot molecules: data needs
	for hot super-earth exoplanets, Mol. Astrophys. 8 (2017) 1--18.
	\newblock \href {http://dx.doi.org/10.1016/j.molap.2017.05.002}
	{\path{doi:10.1016/j.molap.2017.05.002}}.
	
	\bibitem{18IrToGa.H2S}
	P.~G.~J. Irwin, D.~Toledo, R.~Garland, N.~A. Teanby, L.~N. Fletcher, G.~A.
	Orton, B.~Bezard, {Detection of hydrogen sulfide above the clouds in Uranus's
		atmosphere}, Nat. Astron. 2 (2018) 420--427.
	\newblock \href {http://dx.doi.org/10.1038/s41550-018-0432-1}
	{\path{doi:10.1038/s41550-018-0432-1}}.
	
	\bibitem{97Pace.H2S}
	N.~R. Pace, A molecular view of microbial diversity and the biosphere, Science
	276~(5313) (1997) 734--740.
	\newblock \href {http://dx.doi.org/10.1126/science.276.5313.734}
	{\path{doi:10.1126/science.276.5313.734}}.
	
	\bibitem{10KaSaxx}
	L.~Kaltenegger, D.~Sasselov, Detecting planetary geochemical cycles on
	exoplanets: {Atmospheric} signatures and the case of {SO$_2$}, Astrophys. J.
	708 (2010) 1162.
	\newblock \href {http://dx.doi.org/10.1088/0004-637X/708/2/1162}
	{\path{doi:10.1088/0004-637X/708/2/1162}}.
	
	\bibitem{13King.H2S}
	S.~B. King, Potential biological chemistry of hydrogen sulfide ({H$_2$S}) with
	the nitrogen oxides, Free Radical Biology and Medicine 55~(Supplement C)
	(2013) 1 -- 7.
	\newblock \href {http://dx.doi.org/10.1016/j.freeradbiomed.2012.11.005}
	{\path{doi:10.1016/j.freeradbiomed.2012.11.005}}.
	
	\bibitem{17BaAnIv.H2S}
	N.~Bazhanov, M.~Ansar, T.~Ivanciuc, R.~P. Garofalo, A.~Casola, Hydrogen
	sulfide: A novel player in airway development, pathophysiology of respiratory
	diseases, and antiviral defenses, Am. J. Respiratory Cell Molec. Bio. 57
	(2017) 403--410.
	\newblock \href {http://dx.doi.org/10.1165/rcmb.2017-0114TR}
	{\path{doi:10.1165/rcmb.2017-0114TR}}.
	
	\bibitem{08CiCuDe.H2S}
	L.~Ciaffoni, B.~L. Cummings, W.~Denzer, R.~Peverall, S.~R. Procter, G.~A.~D.
	Ritchie, Line strength and collisional broadening studies of hydrogen
	sulphide in the 1.58$\mu$m region using diode laser spectroscopy, Appl. Phys.
	B 92 (2008) 627.
	\newblock \href {http://dx.doi.org/10.1007/s00340-008-3119-y}
	{\path{doi:10.1007/s00340-008-3119-y}}.
	
	\bibitem{72ThKuPe.H2S}
	P.~Thaddeus, M.~L. Kutner, A.~A. Penzias, R.~W. Wilson, K.~B. Jefferts,
	Interstellar hydrogen sulfide, Astron. Astrophys. 176 (1972) L73.
	\newblock \href {http://dx.doi.org/10.1086/181023} {\path{doi:10.1086/181023}}.
	
	\bibitem{jt412}
	T.~Furtenbacher, A.~G. {Cs\'asz\'ar}, J.~Tennyson, {MARVEL: measured active
		rotational-vibrational energy levels}, J. Mol. Spectrosc. 245 (2007)
	115--125.
	\newblock \href {http://dx.doi.org/10.1016/j.jms.2007.07.005}
	{\path{doi:10.1016/j.jms.2007.07.005}}.
	
	\bibitem{12FuCsxx.marvel}
	T.~Furtenbacher, A.~G. {Cs\'asz\'ar}, The role of intensities in determining
	characteristics of spectroscopic networks, J. Molec. Struct. (THEOCHEM) 1009
	(2012) 123 -- 129.
	\newblock \href {http://dx.doi.org/10.1016/j.molstruc.2011.10.057}
	{\path{doi:10.1016/j.molstruc.2011.10.057}}.
	
	\bibitem{12FuCsa}
	T.~Furtenbacher, A.~G. {Cs\'asz\'ar}, {MARVEL: measured active
		rotational-vibrational energy levels. II. Algorithmic improvements}, J.
	Quant. Spectrosc. Radiat. Transf. 113 (2012) 929--935.
	\newblock \href {http://dx.doi.org/10.1016/j.jqsrt.2012.01.005}
	{\path{doi:10.1016/j.jqsrt.2012.01.005}}.
	
	\bibitem{12PoLaVo.H2S}
	E.~R. Polovtseva, N.~A. Lavrentiev, S.~S. Voronina, O.~V. Naumenko, A.~Z.
	Fazliev, Information system for molecular spectroscopy. 5. {Ro-vibrational}
	transitions and energy levels of the hydrogen sulfide molecule, Atmos.
	Oceanic Optics 25 (2012) 157--165.
	\newblock \href {http://dx.doi.org/10.1134/S1024856012020133}
	{\path{doi:10.1134/S1024856012020133}}.
	
	\bibitem{11CsFuxx.marvel}
	A.~G. {Cs\'asz\'ar}, T.~Furtenbacher, Spectroscopic networks, J. Mol.
	Spectrosc. 266 (2011) 99 -- 103.
	\newblock \href {http://dx.doi.org/10.1016/j.jms.2011.03.031}
	{\path{doi:10.1016/j.jms.2011.03.031}}.
	
	\bibitem{16ArPeFu.marvel}
	P.~{\'A}rend{\'a}s, T.~Furtenbacher, A.~G. Cs{\'a}sz{\'a}r, On spectra of
	spectra, J. Math. Chem. 54 (2016) 806--822.
	\newblock \href {http://dx.doi.org/10.1007/s10910-016-0591-1}
	{\path{doi:10.1007/s10910-016-0591-1}}.
	
	\bibitem{jt637}
	T.~Furtenbacher, I.~Szab{\'o}, A.~G. Cs{\'a}sz{\'a}r, P.~F. Bernath, S.~N.
	Yurchenko, J.~Tennyson, Experimental energy levels and partition function of
	the {$^{12}$C$_2$} molecule, Astrophys. J. Suppl. 224 (2016) 44.
	\newblock \href {http://dx.doi.org/10.3847/0067-0049/224/2/44}
	{\path{doi:10.3847/0067-0049/224/2/44}}.
	
	\bibitem{jt672}
	L.~K. McKemmish, T.~Masseron, S.~Sheppard, E.~Sandeman, Z.~Schofield,
	T.~Furtenbacher, A.~G. {Cs\'asz\'ar}, J.~Tennyson, C.~Sousa-Silva, {MARVEL
		analysis of the measured high-resolution spectra of $^{48}$Ti$^{16}$O},
	Astrophys. J. Suppl. 228 (2017) 15.
	\newblock \href {http://dx.doi.org/10.3847/1538-4365/228/2/15}
	{\path{doi:10.3847/1538-4365/228/2/15}}.
	
	\bibitem{jt454}
	J.~Tennyson, P.~F. Bernath, L.~R. Brown, A.~Campargue, M.~R. Carleer, A.~G.
	Cs\'asz\'ar, R.~R. Gamache, J.~T. Hodges, A.~Jenouvrier, O.~V. Naumenko,
	O.~L. Polyansky, L.~S. Rothman, R.~A. Toth, A.~C. Vandaele, N.~F. Zobov,
	L.~Daumont, A.~Z. Fazliev, T.~Furtenbacher, I.~E. Gordon, S.~N. Mikhailenko,
	S.~V. Shirin, {IUPAC critical Evaluation of the Rotational-Vibrational
		Spectra of Water Vapor. Part I. Energy Levels and Transition Wavenumbers for
		H$_2$$^{17}$O and H$_2$$^{18}$O}, J. Quant. Spectrosc. Radiat. Transf. 110
	(2009) 573--596.
	\newblock \href {http://dx.doi.org/10.1016/j.jqsrt.2009.02.014}
	{\path{doi:10.1016/j.jqsrt.2009.02.014}}.
	
	\bibitem{jt482}
	J.~Tennyson, P.~F. Bernath, L.~R. Brown, A.~Campargue, M.~R. Carleer, A.~G.
	Cs\'asz\'ar, L.~Daumont, R.~R. Gamache, J.~T. Hodges, O.~V. Naumenko, O.~L.
	Polyansky, L.~S. Rothman, R.~A. Toth, A.~C. Vandaele, N.~F. Zobov, A.~Z.
	Fazliev, T.~Furtenbacher, I.~E. Gordon, S.~N. Mikhailenko, B.~A. Voronin,
	{IUPAC critical Evaluation of the Rotational-Vibrational Spectra of Water
		Vapor. Part II. Energy Levels and Transition Wavenumbers for HD$^{16}$O,
		HD$^{17}$O, and HD$^{18}$O}, J. Quant. Spectrosc. Radiat. Transf. 111 (2010)
	2160--2184.
	\newblock \href {http://dx.doi.org/10.1016/j.jqsrt.2010.06.012}
	{\path{doi:10.1016/j.jqsrt.2010.06.012}}.
	
	\bibitem{jt539}
	J.~Tennyson, P.~F. Bernath, L.~R. Brown, A.~Campargue, M.~R. Carleer, A.~G.
	Cs\'asz\'ar, L.~Daumont, R.~R. Gamache, J.~T. Hodges, O.~V. Naumenko, O.~L.
	Polyansky, L.~S. Rothmam, A.~C. Vandaele, N.~F. Zobov, A.~R. {Al Derzi},
	C.~F\'abri, A.~Z. Fazliev, T.~Furtenbacher, I.~E. Gordon, L.~Lodi, I.~I.
	Mizus, {IUPAC critical evaluation of the rotational-vibrational spectra of
		water vapor. Part III. Energy levels and transition wavenumbers for
		H$_2$$^{16}$O}, J. Quant. Spectrosc. Radiat. Transf. 117 (2013) 29--80.
	\newblock \href {http://dx.doi.org/10.1016/j.jqsrt.2012.10.002}
	{\path{doi:10.1016/j.jqsrt.2012.10.002}}.
	
	\bibitem{jt576}
	J.~Tennyson, P.~F. Bernath, L.~R. Brown, A.~Campargue, A.~G. Cs\'asz\'ar,
	L.~Daumont, R.~R. Gamache, J.~T. Hodges, O.~V. Naumenko, O.~L. Polyansky,
	L.~S. Rothmam, A.~C. Vandaele, N.~F. Zobov, N.~D\'enes, A.~Z. Fazliev,
	T.~Furtenbacher, I.~E. Gordon, S.-M. Hu, T.~Szidarovszky, I.~A. Vasilenko,
	{IUPAC critical evaluation of the rotational-vibrational spectra of water
		vapor. Part IV. Energy levels and transition wavenumbers for D$_2$$^{16}$O,
		D$_2$$^{17}$O and D$_2$$^{18}$O}, J. Quant. Spectrosc. Radiat. Transf. 142
	(2014) 93--108.
	\newblock \href {http://dx.doi.org/10.1016/j.jqsrt.2014.03.019}
	{\path{doi:10.1016/j.jqsrt.2014.03.019}}.
	
	\bibitem{jtwaterupdate}
	T.~Furtenbacher, J.~Tennyson, O.~V. Naumenko, O.~L. Polyansky, N.~F. Zobov,
	A.~G. Cs\'asz\'ar, {The 2018 Update of the IUPAC Database of Water Energy
		Levels}, J. Quant. Spectrosc. Radiat. Transf.(In preparation).
	
	\bibitem{13FuSzMa.marvel}
	T.~Furtenbacher, T.~Szidarovszky, E.~M{\'a}tyus, C.~F{\'a}bri, A.~G.
	Cs{\'a}sz{\'a}r, {Analysis of the Rotational--Vibrational States of the
		Molecular Ion H$_3^+$}, J. Chem. Theory Comput. 9 (2013) 5471--5478.
	\newblock \href {http://dx.doi.org/10.1021/ct4004355}
	{\path{doi:10.1021/ct4004355}}.
	
	\bibitem{13FuSzFa.marvel}
	T.~Furtenbacher, T.~Szidarovszky, C.~F{\'a}bri, A.~G. Cs{\'a}sz{\'a}r, {MARVEL
		analysis of the rotational--vibrational states of the molecular ions
		H$_2$D$^+$ and D$_2$H$^+$}, Phys. Chem. Chem. Phys. 15 (2013) 10181--10193.
	\newblock \href {http://dx.doi.org/10.1039/c3cp44610g}
	{\path{doi:10.1039/c3cp44610g}}.
	
	\bibitem{jt705}
	K.~L. Chubb, M.~Joseph, J.~Franklin, N.~Choudhury, T.~Furtenbacher, A.~G.
	Cs\'asz\'ar, G.~Gaspard, P.~Oguoko, A.~Kelly, S.~N. Yurchenko, J.~Tennyson,
	C.~Sousa-Silva, {MARVEL analysis of the measured high-resolution spectra of
		C$_2$H$_2$}, J. Quant. Spectrosc. Radiat. Transf. 204 (2018) 42--55.
	\newblock \href {http://dx.doi.org/10.1016/j.jqsrt.2017.08.018}
	{\path{doi:10.1016/j.jqsrt.2017.08.018}}.
	
	\bibitem{jt608}
	A.~R. {Al Derzi}, T.~Furtenbacher, S.~N. Yurchenko, J.~Tennyson, A.~G.
	Cs\'asz\'ar, {MARVEL analysis of the measured high-resolution spectra of
		$^{14}$NH$_3$}, J. Quant. Spectrosc. Radiat. Transf. 161 (2015) 117--130.
	\newblock \href {http://dx.doi.org/10.1016/j.jqsrt.2015.03.034}
	{\path{doi:10.1016/j.jqsrt.2015.03.034}}.
	
	\bibitem{jtNH3update}
	T.~Furtenbacher, P.~A. Coles, J.~Tennyson, A.~G. Cs\'asz\'ar, {Updated MARVEL
		energy levels for ammonia}, J. Quant. Spectrosc. Radiat. Transf.{To be
		submitted}.
	
	\bibitem{11FaMaFu.marvel}
	C.~F{\'a}bri, E.~M{\'a}tyus, T.~Furtenbacher, L.~Nemes, B.~Mih{\'a}ly,
	T.~Zolt{\'a}ni, A.~G. Cs{\'a}sz{\'a}r, Variational quantum mechanical and
	active database approaches to the rotational-vibrational spectroscopy of
	ketene, h$_2$cco, J. Chem. Phys. 135 (2011) 094307.
	
	\bibitem{82HeLixx.C2H2}
	M.~Herman, J.~Lievin, {Acetylene- From intensity alternation in spectra to
		ortho and para molecule}, J. Chem. Educ. 59 (1982) 17.
	\newblock \href {http://dx.doi.org/10.1021/ed059p17}
	{\path{doi:10.1021/ed059p17}}.
	
	\bibitem{56AlPl.H2S}
	H.~C. {Allen Jr.}, E.~K. Plyler, Infrared spectrum of hydrogen sulfide, J.
	Chem. Phys. 25 (1956) 1132--1136.
	\newblock \href {http://dx.doi.org/10.1063/1.1743164}
	{\path{doi:10.1063/1.1743164}}.
	
	\bibitem{98BuJe.method}
	P.~R. Bunker, P.~Jensen, Molecular Symmetry and Spectroscopy, 2nd Edition, NRC
	Research Press, Ottawa, 1998.
	
	\bibitem{jt329}
	A.~Miani, J.~Tennyson, {Can ortho-para transitions for water be observed?}, J.
	Chem. Phys. 120 (2004) 2732--2739.
	
	\bibitem{72HeCoLu.H2S}
	P.~Helminger, R.~L. Cook, F.~C. {De Lucia}, Microwave spectrum and centrifugal
	distortion effects of {{H$_2$S}}, J. Chem. Phys. 56 (1972) 4581--4584.
	\newblock \href {http://dx.doi.org/10.1063/1.1677906}
	{\path{doi:10.1063/1.1677906}}.
	
	\bibitem{95BeYaWi.H2S}
	S.~P. Belov, K.~M.~T. Yamada, G.~Winnewisser, L.~Poteau, R.~Bocquet,
	J.~Demaison, O.~Polyansky, M.~Y. Tretyakov, Terahertz rotational spectrum of
	{H$_2$S}, J. Mol. Spectrosc. 173 (1995) 380--390.
	\newblock \href {http://dx.doi.org/10.1006/jmsp.1995.1242}
	{\path{doi:10.1006/jmsp.1995.1242}}.
	
	\bibitem{68CuKeGa.H2S}
	R.~E. Cupp, R.~A. Keikpf, J.~J. Gallagher, Hyperfine structure in the
	millimeter spectrum of hydrogen sulfide electric spectroscopy on
	asymmetric-top molecules, Phys. Rev. 171 (1968) 60--69.
	\newblock \href {http://dx.doi.org/10.1103/PhysRev.171.60}
	{\path{doi:10.1103/PhysRev.171.60}}.
	
	\bibitem{71Huiszoon.H2S}
	C.~Huiszoon, A high resolution spectrometer for the shorter millimeter
	wavelength region, Rev. Sci. Instrum. 42 (1971) 477--481.
	\newblock \href {http://dx.doi.org/10.1063/1.1685135}
	{\path{doi:10.1063/1.1685135}}.
	
	\bibitem{14CaPu.H2S}
	G.~Cazzoli, C.~Puzzarini, The rotational spectrum of hydrogen sulfide: The
	{H$_2$$^{33}$S} and {H$_2$$^{32}$S} isotopologues revisited, J. Mol.
	Spectrosc. 298~(Supplement C) (2014) 31 -- 37.
	\newblock \href {http://dx.doi.org/10.1016/j.jms.2014.02.002}
	{\path{doi:10.1016/j.jms.2014.02.002}}.
	
	\bibitem{85BuFeMe.H2S}
	A.~V. Burenin, T.~M. Fevralskikh, A.~A. Melnikov, S.~M. Shapin, Microwave
	spectrum of the hydrogen sulfide molecule {H$_2$$^{32}$S} in the ground
	state, J. Mol. Spectrosc. 109 (1985) 1--7.
	\newblock \href {http://dx.doi.org/10.1016/0022-2852(85)90045-1}
	{\path{doi:10.1016/0022-2852(85)90045-1}}.
	
	\bibitem{94YaKlxx.H2S}
	K.~M.~T. Yamada, S.~Klee, Pure rotational spectrum of {H$_{2}$S} in the
	far-infrared region measured by {FTIR} spectroscopy, J. Mol. Spectrosc. 166
	(1994) 395--405.
	\newblock \href {http://dx.doi.org/10.1006/jmsp.1994.1204}
	{\path{doi:10.1006/jmsp.1994.1204}}.
	
	\bibitem{13CaPu.H2S}
	G.~Cazzoli, C.~Puzzarini, Sub-doppler resolution in the {THz} frequency domain:
	1 {kHz} accuracy at 1 {THz} by exploiting the lamb-dip technique, J. Phys.
	Chem. A 117 (2013) 13759--13766.
	\newblock \href {http://dx.doi.org/10.1021/jp407980f}
	{\path{doi:10.1021/jp407980f}}.
	
	\bibitem{jt558}
	A.~A.~A. Azzam, S.~N. Yurchenko, J.~Tennyson, M.-A. Martin, O.~Pirali,
	{Terahertz spectroscopy of hydrogen sulfide}, J. Quant. Spectrosc. Radiat.
	Transf. 130 (2013) 341--51.
	\newblock \href {http://dx.doi.org/10.1016/j.jqsrt.2013.05.035}
	{\path{doi:10.1016/j.jqsrt.2013.05.035}}.
	
	\bibitem{83FlCaJo.H2S}
	J.-M. Flaud, C.~Camy-Peyret, J.~W.~C. Johns, The far-infrared spectrum of
	hydrogen-sulfide - the (000) rotational-constants of {H$_2$$^{32}$S},
	{H$_2$$^{33}$S} and {H$_2$$^{34}$S}, Can. J. Phys. 61 (1983) 1462--1473.
	\newblock \href {http://dx.doi.org/10.1139/p83-188}
	{\path{doi:10.1139/p83-188}}.
	
	\bibitem{18UlBeGr.H2S}
	O.~N. Ulenikov, E.~S. Bekhtereva, O.~V. Gromova, P.~A. Glushkov, A.~P.
	Scherbakov, V.-M. Horneman, C.~Sydow, C.~Maul, S.~Bauerecker, Extended
	analysis of the high resolution {FTIR} spectra of {H$_2^M$S} (m =
	32,33,34,36) in the region of the bending fundamental band: the $\nu_2$ and
	bands: line positions, strengths, and pressure broadening widths, J. Quant.
	Spectrosc. Radiat. Transf. 216 (2018) 76 -- 98.
	\newblock \href {http://dx.doi.org/10.1016/j.jqsrt.2018.05.009}
	{\path{doi:10.1016/j.jqsrt.2018.05.009}}.
	
	\bibitem{82LaEdGi.H2S}
	W.~C. Lane, T.~H. Edwards, J.~R. Gillis, F.~S. Bonomo, F.~J. Murcray, Analysis
	of $\nu_{2}$ of {H$_{2}$S}, J. Mol. Spectrosc. 95 (1982) 365--380.
	\newblock \href {http://dx.doi.org/10.1016/0022-2852(82)90136-9}
	{\path{doi:10.1016/0022-2852(82)90136-9}}.
	
	\bibitem{83Strow.H2S}
	L.~L. Strow, Measurement and analysis of the $\nu_{2}$ band of {H$_{2}$S}:
	Comparison among several reduced forms of the rotational hamiltonian, J. Mol.
	Spectrosc. 97 (1983) 9--28.
	\newblock \href {http://dx.doi.org/10.1016/0022-2852(83)90334-X}
	{\path{doi:10.1016/0022-2852(83)90334-X}}.
	
	\bibitem{96UlMaKo.H2S}
	O.~N. Ulenikov, A.~B. Malikova, M.~Koivusaari, S.~Alanko, R.~Anttila, High
	resolution vibrational rotational spectrum of {H$_2$S} in the region of the
	$\nu_2$ fundamental band, J. Mol. Spectrosc. 176 (1996) 229--235.
	\newblock \href {http://dx.doi.org/10.1006/jmsp.1996.0082}
	{\path{doi:10.1006/jmsp.1996.0082}}.
	
	\bibitem{98BrCrCr.H2S}
	L.~R. Brown, J.~A. Crisp, D.~Crisp, O.~V. Naumenko, M.~A. Smirnov, L.~N.
	Sinitsa, A.~Perrin, The absorption spectrum of {H$_2$S} between 2150 and 4260
	cm$^{-1}$: Analysis of the positions and intensities in the first
	($2\nu_{2}$, $\nu_{1}$, and $\nu_{3}$) and second ($3\nu_{2}$, $\nu_{1} /
	\nu_{2}$, and $\nu_{2} / \nu_{3}$) triad regions, J. Mol. Spectrosc. 188
	(1998) 148--174.
	\newblock \href {http://dx.doi.org/10.1006/jmsp.1997.7501}
	{\path{doi:10.1006/jmsp.1997.7501}}.
	
	\bibitem{17Horneman.H2S}
	V.~M. Horneman, Private Communication.
	
	\bibitem{84LeFlCa.H2S}
	L.~Lechuga-Fossat, J.-M. Flaud, C.~Camy-Peyret, J.~W.~C. Johns, The spectrum of
	natural hydrogen-sulfide between 2150 cm$^{-1}$ and 2950 cm$^{-1}$, Can. J.
	Phys. {62} ({1984}) 1889--1923.
	\newblock \href {http://dx.doi.org/10.1139/p84-233}
	{\path{doi:10.1139/p84-233}}.
	
	\bibitem{81GiEd.H2S}
	J.~R. Gillis, T.~H. Edwards, {Analysis of $2\nu_2$, $\nu_1$, and $\nu_3$ of
		H$_{2}$S}, J. Mol. Spectrosc. 85 (1981) 55--73.
	\newblock \href {http://dx.doi.org/10.1016/0022-2852(81)90309-X}
	{\path{doi:10.1016/0022-2852(81)90309-X}}.
	
	\bibitem{96UlOnKo.H2S}
	O.~N. Ulenikov, G.~A. Onopenko, M.~Koivusaari, S.~Alanko, R.~Anttila, High
	resolution fourier transform spectrum of {H$_2$S} in the 3300-4080 cm$^{-1}$
	region, J. Mol. Spectrosc. 176 (1996) 236--250.
	\newblock \href {http://dx.doi.org/10.1006/jmsp.1996.0083}
	{\path{doi:10.1006/jmsp.1996.0083}}.
	
	\bibitem{05UlLiBe.H2S}
	O.~N. Ulenikov, A.-W. Liu, E.~S. Bekhtereva, O.~V. Gromova, L.-Y. Hao, S.-M.
	Hu, High-resolution fourier transform spectrum of {H$_2$S} in the region of
	the second hexade, J. Mol. Spectrosc. 234 (2005) 270--278.
	\newblock \href {http://dx.doi.org/10.1016/j.jms.2005.09.010}
	{\path{doi:10.1016/j.jms.2005.09.010}}.
	
	\bibitem{97BrCrCr.H2S}
	L.~R. Brown, J.~A. Crisp, D.~Crisp, O.~V. Naumenko, M.~A. Smirnov, L.~N.
	Sinitsa, The first hexad of interacting states of {H$_{2}$S} molecule, SPIE
	3090 (1997) 111--113.
	\newblock \href {http://dx.doi.org/10.1117/12.267745}
	{\path{doi:10.1117/12.267745}}.
	
	\bibitem{18Liu.H2S}
	A.-W. Liu, Private Communication.
	
	\bibitem{04BrNaPoa.H2S}
	L.~R. Brown, O.~V. Naumenko, E.~R. Polovtseva, L.~N. Sinitsa, Hydrogen sulfide
	absorption spectrum in the 5700--6600 cm$^{-1}$ spectral region, Proceedings
	14th Symposium on High-Resolution Molecular Spectroscopy 5311 (2004) 59--67.
	\newblock \href {http://dx.doi.org/10.1117/12.545192}
	{\path{doi:10.1117/12.545192}}.
	
	\bibitem{04BrNaPoSi.H2S}
	L.~R. Brown, O.~V. Naumenko, E.~R. Polovtseva, L.~N. Sinitsa, Absorption
	spectrum of {H$_2$S} between 7200 and 7890 cm$^{-1}$, Proc. SPIE 5396 (2004)
	5396--5397.
	\newblock \href {http://dx.doi.org/10.1117/12.548211}
	{\path{doi:10.1117/12.548211}}.
	
	\bibitem{04UlLiBe.H2S}
	O.~N. Ulenikov, A.-W. Liu, E.~S. Bekhtereva, O.~V. Gromova, L.-Y. Hao, S.-M.
	Hu, On the study of high-resolution rovibrational spectrum of {H$_2$S} in the
	region of 7300--7900 cm$^{-1}$, J. Mol. Spectrosc. 226 (2004) 57--70.
	\newblock \href {http://dx.doi.org/10.1016/j.jms.2004.03.014}
	{\path{doi:10.1016/j.jms.2004.03.014}}.
	
	\bibitem{04UlLiBeb.H2S}
	O.~N. Ulenikov, A.-W. Liu, E.~S. Bekhtereva, S.~V. Grebneva, W.-P. Deng, O.~V.
	Gromova, S.-M. Hu, High resolution fourier transform spectrum of {H$_2$S} in
	the region of 8500-8900 cm$^{-1}$, J. Mol. Spectrosc. 228 (2004) 110--119.
	\newblock \href {http://dx.doi.org/10.1016/j.jms.2004.07.011}
	{\path{doi:10.1016/j.jms.2004.07.011}}.
	
	\bibitem{04BrNaPob.H2S}
	L.~R. Brown, O.~V. Naumenko, E.~R. Polovtseva, L.~N. Sinitsa, Hydrogen sulfide
	absorption spectrum in the 8400--8900 cm$^{-1}$ spectral region, Eleventh
	International Symposium on Atmospheric and Ocean Optics/Atmospheric Physics
	5743 (2004) 1--7.
	\newblock \href {http://dx.doi.org/10.1117/12.606253}
	{\path{doi:10.1117/12.606253}}.
	
	\bibitem{03DiNaHu.H2S}
	Y.~Ding, O.~Naumenko, S.-M. Hu, Q.~Zhu, E.~Bertseva, A.~Campargue, {The
		absorption spectrum of H$_2$S between 9540 and 10000 cm$^{-1}$ by intracavity
		laser absorption spectroscopy with a vertical external cavity surface
		emitting laser}, J. Mol. Spectrosc. 217 (2003) 222--238.
	\newblock \href {http://dx.doi.org/10.1016/S0022-2852(02)00037-1}
	{\path{doi:10.1016/S0022-2852(02)00037-1}}.
	
	\bibitem{01NaCaxxa.H2S}
	O.~Naumenko, A.~Campargue, {H$_2$$^{32}$S : First Observation of the
		(70$\pm$,0) Local Mode Pair and Updated Global Effective Vibrational
		Hamiltonian}, J. Mol. Spectrosc. 210 (2001) 224--232.
	\newblock \href {http://dx.doi.org/10.1006/jmsp.2001.8460}
	{\path{doi:10.1006/jmsp.2001.8460}}.
	
	\bibitem{94GrRaSt.H2S}
	R.~Gro{\ss}klo{\ss}, S.~B. Rai, R.~Stuber, W.~Demtroder, Diode laser overtone
	spectroscopy of hydrogen sulfide, Chem. Phys. Lett. 229 (1994) 609--615.
	\newblock \href {http://dx.doi.org/10.1016/0009-2614(94)01079-X}
	{\path{doi:10.1016/0009-2614(94)01079-X}}.
	
	\bibitem{97VaBiCa.H2S}
	O.~Vaittinen, L.~Biennier, A.~Campargue, J.-M. Flaud, L.~Halonen, Local mode
	effects on the high-resolution overtone spectrum of {H$_2$S} around 12500
	cm$^{-1}$, J. Mol. Spectrosc. 184 (1997) 288--299.
	\newblock \href {http://dx.doi.org/10.1006/jmsp.1997.7319}
	{\path{doi:10.1006/jmsp.1997.7319}}.
	
	\bibitem{99CaFlxx.H2S}
	A.~Campargue, J.-M. Flaud, The overtone spectrum of {H$_2$$^{32}$S} near 13 200
	cm$^{-1}$, J. Mol. Spectrosc. 194 (1999) 43--51.
	\newblock \href {http://dx.doi.org/10.1006/jmsp.1998.7754}
	{\path{doi:10.1006/jmsp.1998.7754}}.
	
	\bibitem{01NaCaxxb.H2S}
	O.~Naumenko, A.~Campargue, Local mode effects in the absorption spectrum of
	{H$_2$S} between 10780 and 11330 cm$^{-1}$, J. Mol. Spectrosc. 209 (2001)
	242--253.
	\newblock \href {http://dx.doi.org/10.1006/jmsp.2001.8417}
	{\path{doi:10.1006/jmsp.2001.8417}}.
	
	\bibitem{02CoRoTy.H2S}
	T.~Cours, P.~Rosmus, V.~G. Tyuterev, Ab initio dipole moment functions of
	{H$_2$$^{32}$S} and intensity anomalies in rovibrational spectra, J. Chem.
	Phys. 117 (2002) 5192--5208.
	\newblock \href {http://dx.doi.org/10.1063/1.1499487}
	{\path{doi:10.1063/1.1499487}}.
	
	\bibitem{94WaKuSu.H2S}
	J.~Waschull, F.~Kuhnemann, B.~Sumpf, Self-, air, and helium broadening in the
	$\nu_2$ band of {H$_{2}$S}, J. Mol. Spectrosc. 165 (1994) 150--158.
	\newblock \href {http://dx.doi.org/10.1006/jmsp.1994.1117}
	{\path{doi:10.1006/jmsp.1994.1117}}.
	
	\bibitem{96SuMeKr.H2S}
	B.~Sumpf, I.~Meusel, H.-D. Kronfeldt, Self- and air-broadening in the $\nu_{1}$
	and $\nu_{3}$ bands of {H$_2$S}, J. Mol. Spectrosc. 177 (1996) 143--145.
	\newblock \href {http://dx.doi.org/10.1006/jmsp.1996.0126}
	{\path{doi:10.1006/jmsp.1996.0126}}.
	
	\bibitem{97SuMeKr.H2S}
	B.~Sumpf, I.~Meusel, H.-D. Kronfeldt, Noble gas broadening in fundamental bands
	of {H$_2$S}, J. Mol. Spectrosc. 184 (1997) 51--55.
	\newblock \href {http://dx.doi.org/10.1006/jmsp.1997.7290}
	{\path{doi:10.1006/jmsp.1997.7290}}.
	
	\bibitem{97Suxxxx.H2S}
	B.~Sumpf, {Experimental Investigation of the Self-Broadening Coefficients in
		the $\nu_{1}$ / $\nu_{3}$ Band of SO$_2$ and the $2\nu_{2}$ Band of H$_2$S},
	J. Mol. Spectrosc. 181 (1997) 160--167.
	\newblock \href {http://dx.doi.org/10.1006/jmsp.1996.7168}
	{\path{doi:10.1006/jmsp.1996.7168}}.
	
	\bibitem{98PiPoCo}
	H.~M. Pickett, R.~L. Poynter, E.~A. Cohen, M.~L. Delitsky, J.~C. Pearson,
	H.~S.~P. Muller, Submillimeter, millimeter, and microwave spectral line
	catalog, J. Quant. Spectrosc. Radiat. Transf. 60 (1998) 883--890.
	\newblock \href {http://dx.doi.org/10.1016/S0022-4073(98)00091-0}
	{\path{doi:10.1016/S0022-4073(98)00091-0}}.
	
	\bibitem{73HeDeKi.H2S}
	P.~Helminger, F.~C. {De Lucia}, W.~H. Kirchhoff, Microwave spectra of molecules
	of astrophysical interest {IV}. {Hydrogen} sulfide, J. Phys. Chem. Ref. Data
	2 (1973) 215--223.
	\newblock \href {http://dx.doi.org/10.1063/1.3253117}
	{\path{doi:10.1063/1.3253117}}.
	
	\bibitem{02KiSuKr.H2S}
	A.~Kissel, B.~Sumpf, H.-D. Kronfeldt, B.~A. Tikhomirov, Y.~N. Ponomarev,
	Molecular-gas-pressure-induced line-shift and line-broadening in the
	$\nu_2$-band of {H$_2$S}, J. Mol. Spectrosc. 216 (2002) 345--354.
	\newblock \href {http://dx.doi.org/10.1006/jmsp.2002.8630}
	{\path{doi:10.1006/jmsp.2002.8630}}.
	
	\bibitem{98RoRiGo.db}
	L.~S. Rothman, C.~P. Rinsland, A.~Goldman, S.~T. Massie, D.~P. Edwards, J.~M.
	Flaud, A.~Perrin, C.~Camy-Peyret, V.~Dana, J.~Y. Mandin, J.~Schroeder,
	A.~McCann, R.~R. Gamache, R.~Wattson, K.~Yoshino, K.~V. Chance, K.~W. Jucks,
	L.~R. Brown, V.~Nemtchinov, P.~Varanasi, The {HITRAN} molecular spectroscopic
	database and {HAWKS} ({HITRAN} atmospheric workstation): 1996 edition, J.
	Quant. Spectrosc. Radiat. Transf. 60 (1998) 665--710.
	\newblock \href {http://dx.doi.org/10.1016/S0022-4073(98)00078-8}
	{\path{doi:10.1016/S0022-4073(98)00078-8}}.
	
	\bibitem{87LeFlCa.H2S}
	L.~Lechuga-Fossat, J.-M. Flaud, C.~Camy-Peyret, P.~Arcas, M.~Cuisenier, The
	{H$_2$S} spectrum in the 1.6$\mu$m spectral region, Mol. Phys. 61 (1987)
	23--32.
	\newblock \href {http://dx.doi.org/10.1080/00268978700100961}
	{\path{doi:10.1080/00268978700100961}}.
	
	\bibitem{13Azzamx.H2S}
	A.~A.~A. Azzam, {A linelist for the hydrogen sulphide molecule}, Ph.D. thesis,
	University College London, the thesis is available on-line at {\footnotesize
		\texttt{http://discovery.ucl.ac.uk/1404058/}} (2013).
	
	\bibitem{94KoJexx.H2S}
	I.~N. Kozin, P.~Jensen, Fourfold clusters of rotational energy levels for
	{H$_{2}$S} studied with a potential energy surface derived from experiment,
	J. Mol. Spectrosc. 163 (1994) 483--509.
	\newblock \href {http://dx.doi.org/10.1006/jmsp.1994.1041}
	{\path{doi:10.1006/jmsp.1994.1041}}.
	
	\bibitem{95FlGrRa.H2S}
	J.~M. Flaud, R.~Grosskloss, S.~B. Rai, R.~Struber, W.~Demtroder, D.~A. Tate,
	L.~guo Wang, T.~F. Gallagher, {Diode laser Spectroscopy of $H_{2}^{32}S$
		around 0.82 $\mu$m}, J. Mol. Spectrosc. 172 (1995) 275--281.
	\newblock \href {http://dx.doi.org/10.1006/jmsp.1995.1175}
	{\path{doi:10.1006/jmsp.1995.1175}}.
	
	\bibitem{85LaEdGi.H2S}
	W.~C. Lane, T.~H. Edwards, J.~R. Gillis, F.~S. Bonomo, F.~J. Murcray, Analysis
	of $\nu_{2}$ of {H$_{2}$$^{33}$S} and {H$_{2}$$^{34}$S}, J. Mol. Spectrosc.
	111 (1985) 320--326.
	\newblock \href {http://dx.doi.org/10.1016/0022-2852(85)90008-6}
	{\path{doi:10.1016/0022-2852(85)90008-6}}.
	
	\bibitem{94ByNaSm.H2S}
	A.~D. Bykov, O.~V. Naumenko, M.~A. Smirnov, L.~N. Sinitsa, L.~R. Brown,
	J.~Crisp, D.~Crisp, The infrared-spectrum of {H$_2$S} from 1 to 5 $\mu$m,
	Can. J. Phys. {72} ({1994}) 989--1000.
	\newblock \href {http://dx.doi.org/10.1139/p94-130}
	{\path{doi:10.1139/p94-130}}.
	
	\bibitem{01TyTaSc.H2S}
	V.~G. Tyuterev, S.~A. Tashkun, D.~W. Schwenke, An accurate isotopically
	invariant potential function of the hydrogen sulphide molecule, Chem. Phys.
	Lett. 348 (2001) 223--234.
	\newblock \href {http://dx.doi.org/10.1016/S0009-2614(01)01093-4}
	{\path{doi:10.1016/S0009-2614(01)01093-4}}.
	
	\bibitem{69MiLeHa.H2S}
	R.~E. Miller, G.~E. Leroi, T.~M. Hard, Analysis of the pure rotational
	absorption spectra of hydrogen sulfide and deuterium sulfide, J. Chem. Phys.
	50 (1969) 677--684.
	\newblock \href {http://dx.doi.org/10.1063/1.1671116}
	{\path{doi:10.1063/1.1671116}}.
	
	\bibitem{69SnEd.H2S}
	L.~E. Snyder, T.~H. Edwards, Simultaneous analysis of the (110) and (011) bands
	of hydrogen sulfide, J. Mol. Spectrosc. 31 (1969) 347--361.
	\newblock \href {http://dx.doi.org/10.1016/0022-2852(69)90365-8}
	{\path{doi:10.1016/0022-2852(69)90365-8}}.
	
	\bibitem{jt557}
	L.~S. Rothman, I.~E. Gordon, Y.~Babikov, A.~Barbe, D.~C. Benner, P.~F. Bernath,
	M.~Birk, L.~Bizzocchi, V.~Boudon, L.~R. Brown, A.~Campargue, K.~Chance, E.~A.
	Cohen, L.~H. Coudert, V.~M. Devi, B.~J. Drouin, A.~Fayt, J.-M. Flaud, R.~R.
	Gamache, J.~J. Harrison, J.-M. Hartmann, C.~Hill, J.~T. Hodges,
	D.~Jacquemart, A.~Jolly, J.~Lamouroux, R.~J. {Le Roy}, G.~Li, D.~A. Long,
	O.~M. Lyulin, C.~J. Mackie, S.~T. Massie, S.~Mikhailenko, H.~S.~P.
	M{\"u}ller, O.~V. Naumenko, A.~V. Nikitin, J.~Orphal, V.~Perevalov,
	A.~Perrin, E.~R. Polovtseva, C.~Richard, M.~A.~H. Smith, E.~Starikova,
	K.~Sung, S.~Tashkun, J.~Tennyson, G.~C. Toon, V.~G. Tyuterev, G.~Wagner, {The
		{\it HITRAN} 2012 molecular spectroscopic database}, J. Quant. Spectrosc.
	Radiat. Transf. 130 (2013) 4 -- 50.
	\newblock \href {http://dx.doi.org/10.1016/jqsrt.2013.07.002}
	{\path{doi:10.1016/jqsrt.2013.07.002}}.
	
	\bibitem{jt691}
	I.~E. Gordon, L.~S. Rothman, C.~Hill, R.~V. Kochanov, Y.~Tan, P.~F. Bernath,
	M.~Birk, V.~Boudon, A.~Campargue, K.~V. Chance, B.~J. Drouin, J.-M. Flaud,
	R.~R. Gamache, J.~T. Hodges, D.~Jacquemart, V.~I. Perevalov, A.~Perrin, K.~P.
	Shine, M.-A.~H. Smith, J.~Tennyson, G.~C. Toon, H.~Tran, V.~G. Tyuterev,
	A.~Barbe, A.~G. Cs{\'a}sz{\'a}r, V.~M. Devi, T.~Furtenbacher, J.~J. Harrison,
	J.-M. Hartmann, A.~Jolly, T.~J. Johnson, T.~Karman, I.~Kleiner, A.~A.
	Kyuberis, J.~Loos, O.~M. Lyulin, S.~T. Massie, S.~N. Mikhailenko,
	N.~Moazzen-Ahmadi, H.~S.~P. M{\"u}ller, O.~V. Naumenko, A.~V. Nikitin, O.~L.
	Polyansky, M.~Rey, M.~Rotger, S.~W. Sharpe, K.~Sung, E.~Starikova, S.~A.
	Tashkun, J.~{Vander Auwera}, G.~Wagner, J.~Wilzewski, P.~Wcis{\l}o, S.~Yu,
	E.~J. Zak, {The {\it HITRAN} 2016 molecular spectroscopic database}, J.
	Quant. Spectrosc. Radiat. Transf. 203 (2017) 3--69.
	\newblock \href {http://dx.doi.org/10.1016/j.jqsrt.2017.06.038}
	{\path{doi:10.1016/j.jqsrt.2017.06.038}}.
	
	\bibitem{92RoGaTi.db}
	L.~Rothman, R.~R. Gamache, R.~H. Tipping, C.~P. Rinsland, M.~A.~H. Smith, D.~C.
	Benner, V.~M. Devi, J.~M. Flaud, C.~Camy-Peyret, A.~Perrin, A.~Goldman, S.~T.
	Massie, L.~R. Brown, R.~A. Toth, The {HITRAN} molecular database - editions
	of 1991 and 1992, J. Quant. Spectrosc. Radiat. Transf. 48 (1992) 469--507.
	\newblock \href {http://dx.doi.org/10.1016/0022-4073(92)90115-K}
	{\path{doi:10.1016/0022-4073(92)90115-K}}.
	
	\bibitem{01RoBaBe.db}
	L.~S. Rothman, A.~Barbe, D.~C. Benner, L.~R. Brown, C.~Camy-Peyret, M.~R.
	Carleer, K.~Chance, C.~Clerbaux, V.~Dana, V.~M. Devi, A.~Fayt, J.~M. Flaud,
	R.~R. Gamache, A.~Goldman, D.~Jacquemart, K.~W. Jucks, W.~J. Lafferty, J.~Y.
	Mandin, S.~T. Massie, V.~Nemtchinov, D.~A. Newnham, A.~Perrin, C.~P.
	Rinsland, J.~Schroeder, K.~M. Smith, M.~A.~H. Smith, K.~Tang, R.~A. Toth,
	J.~Vander~Auwera, P.~Varanasi, K.~Yoshino, The {HITRAN} molecular
	spectroscopic database: edition of 2000 including updates through 2001, J.
	Quant. Spectrosc. Radiat. Transf. 82 (2003) 5--44.
	\newblock \href {http://dx.doi.org/10.1016/S0022-4073(03)00146-8}
	{\path{doi:10.1016/S0022-4073(03)00146-8}}.
	
	\bibitem{jt350}
	L.~S. Rothman, D.~Jacquemart, A.~Barbe, D.~C. Benner, M.~Birk, L.~R. Brown,
	M.~R. Carleer, C.~Chackerian, K.~Chance, L.~H. Coudert, V.~Dana, V.~M. Devi,
	J.-M. Flaud, R.~R. Gamache, A.~Goldman, J.-M. Hartmann, K.~W. Jucks, A.~G.
	Maki, J.-Y. Mandin, S.~T. Massie, J.~Orphal, A.~Perrin, C.~P. Rinsland,
	M.~A.~H. Smith, J.~Tennyson, R.~N. Tolchenov, R.~A. Toth, J.~Vander~Auwera,
	P.~Varanasi, G.~Wagner, {The {\it HITRAN} 2004 molecular spectroscopic
		database}, J. Quant. Spectrosc. Radiat. Transf. 96 (2005) 139--204.
	
	\bibitem{jt453}
	L.~S. Rothman, I.~E. Gordon, A.~Barbe, D.~C. Benner, P.~F. Bernath, M.~Birk,
	V.~Boudon, L.~R. Brown, A.~Campargue, J.~P. Champion, K.~Chance, L.~H.
	Coudert, V.~Dana, V.~M. Devi, S.~Fally, J.~M. Flaud, R.~R. Gamache,
	A.~Goldman, D.~Jacquemart, I.~Kleiner, N.~Lacome, W.~J. Lafferty, J.~Y.
	Mandin, S.~T. Massie, S.~N. Mikhailenko, C.~E. Miller, N.~Moazzen-Ahmadi,
	O.~V. Naumenko, A.~V. Nikitin, J.~Orphal, V.~I. Perevalov, A.~Perrin,
	A.~Predoi-Cross, C.~P. Rinsland, M.~Rotger, M.~Simeckova, M.~A.~H. Smith,
	K.~Sung, S.~A. Tashkun, J.~Tennyson, R.~A. Toth, A.~C. Vandaele,
	J.~Vander~Auwera, {The {\it HITRAN} 2008 molecular spectroscopic database},
	J. Quant. Spectrosc. Radiat. Transf. 110 (2009) 533--572.
	
	\bibitem{jt640}
	A.~A.~A. Azzam, S.~N. Yurchenko, J.~Tennyson, O.~V. Naumenko, {ExoMol line
		lists XVI: A Hot Line List for H$_{2}$S}, Mon. Not. R. Astron. Soc. 460
	(2016) 4063--4074.
	\newblock \href {http://dx.doi.org/10.1093/mnras/stw1133}
	{\path{doi:10.1093/mnras/stw1133}}.
	
	\bibitem{jt528}
	J.~Tennyson, S.~N. Yurchenko, {ExoMol: molecular line lists for exoplanet and
		other atmospheres}, Mon. Not. R. Astron. Soc. 425 (2012) 21--33.
	\newblock \href {http://dx.doi.org/10.1111/j.1365-2966.2012.21440.x}
	{\path{doi:10.1111/j.1365-2966.2012.21440.x}}.
	
	\bibitem{jt631}
	J.~Tennyson, S.~N. Yurchenko, A.~F. Al-Refaie, E.~J. Barton, K.~L. Chubb, P.~A.
	Coles, S.~Diamantopoulou, M.~N. Gorman, C.~Hill, A.~Z. Lam, L.~Lodi, L.~K.
	McKemmish, Y.~Na, A.~Owens, O.~L. Polyansky, T.~Rivlin, C.~Sousa-Silva, D.~S.
	Underwood, A.~Yachmenev, E.~Zak, {The ExoMol database: molecular line lists
		for exoplanet and other hot atmospheres}, J. Mol. Spectrosc. 327 (2016)
	73--94.
	\newblock \href {http://dx.doi.org/10.1016/j.jms.2016.05.002}
	{\path{doi:10.1016/j.jms.2016.05.002}}.
	
	\bibitem{jt570}
	R.~J. Barber, J.~K. Strange, C.~Hill, O.~L. Polyansky, G.~C. Mellau, S.~N.
	Yurchenko, J.~Tennyson, {ExoMol line lists -- III. An improved hot
		rotation-vibration line list for HCN and HNC}, Mon. Not. R. Astron. Soc. 437
	(2014) 1828--1835.
	\newblock \href {http://dx.doi.org/10.1093/mnras/stt2011}
	{\path{doi:10.1093/mnras/stt2011}}.
	
	\bibitem{17McChRi}
	L.~K. McKemmish, K.~L. Chubb, T.~Rivlin, J.~S. Baker, M.~N. Gorman, A.~Heward,
	W.~Dunn, M.~Tessenyi, {Bringing pupils into the ORBYTS of research},
	Astronomy \& Geophysics 58~(5) (2017) 5.11.
	\newblock \href {http://dx.doi.org/10.1093/astrogeo/atx169}
	{\path{doi:10.1093/astrogeo/atx169}}.
	
	\bibitem{jt709}
	C.~Sousa-Silva, L.~K. McKemmish, K.~L. Chubb, J.~Baker, E.~J. Barton, M.~N.
	Gorman, T.~Rivlin, J.~Tennyson, {Original Research By Young Twinkle Students
		(ORBYTS): When can students start performing original research?}, Phys. Educ.
	53 (2018) 015020.
	\newblock \href {http://dx.doi.org/10.1088/1361-6552/aa8f2a}
	{\path{doi:10.1088/1361-6552/aa8f2a}}.
	
\end{thebibliography}

\end{document}